\documentclass[manuscript]{aastex}

\slugcomment{Accepted by the Astrophysical Journal}

\shorttitle{Selective desorption in interstellar ice formation}
\shortauthors{Kalv\=ans}

\begin{document}

\title{The effect of selective desorption mechanisms during interstellar ice formation}

\author{J. Kalv\=ans}
\affil{Engineering Research Institute "Ventspils International Radio Astronomy Center" of Ventspils University College,
    Inzenieru 101, Ventspils, Latvia, LV-3601}
\email{juris.kalvans@venta.lv}

\begin{abstract}
Major components of ices on interstellar grains in molecular clouds -- water and carbon oxides -- occur at various optical depths. This implies that selective desorption mechanisms are at work. An astrochemical model of a contracting low-mass molecular cloud core is presented. Ice was treated as consisting of the surface and three subsurface layers (i.e., sublayers). Photodesorption, reactive desorption, and indirect reactive desorption were investigated. The latter manifests itself through desorption from H+H reaction on grains. Desorption of shallow subsurface species was also included. Modeling results suggest the existence of a "photon-dominated ice" during the early phases of core contraction. Subsurface ice is chemically processed by interstellar photons, which produces complex organic molecules (COMs). Desorption from the subsurface layer results in high COM gas-phase abundances at $A_V=2.4--10$mag. This may contribute toward an explanation for COM observations in dark cores. It was found that photodesorption mostly governs the onset of ice accumulation onto grains. Reaction-specific reactive desorption is efficient for small molecules that form via highly exothermic atom-addition reactions. Higher reactive desorption efficiency results in lower gas-phase abundances of COMs. Indirect reactive desorption allows for closely reproducing the observed H$_2$O:CO:CO$_2$ ratio toward a number of background stars. Presumably, this can be done by any mechanism whose efficiency fits with the sequence $\rm CO\geq CO_2>>H_2O$. After the freeze-out has ended, the three sublayers represent chemically distinct parts of the mantle. The likely $A_V$ threshold for the appearance of CO ice is 8--10.5mag . The lower value is supported by observations.
\end{abstract}


\section{Introduction}
\label{intro}

Interstellar ices occur on the surface of dust grains in dark interstellar clouds. Cold and dense molecular cores form in the clouds via gravitational contraction. The cores evolve into protostars over a time-scale of order 1Myr \citep{Lee99}. During this evolution, molecules settle onto the dust-grains. In dark cores, almost a complete freeze-out occurs. An ice mantle, probably hundreds of monolayers (ML) thick, accumulates onto a typical grain. Surface reactions are a major route for molecule formation \citep{Pickles77} in interstellar ices. Along with accretion, this process defines the initial composition for all subsurface ice layers. Gas and surface chemistry are influenced by thermal and non-thermal desorption mechanisms. Selective desorption mechanisms, whose efficiency depends on surface species' adsorption energy $E_D$, is of particular importance because of its ability to affect the relative proportions of abundances for solid species. Non-selective mechanisms also may have an effect on these proportions by determining the starting time and length of the ice formation epoch. These temporal parameters, in turn, affect the physical conditions at which the formation of ice occurs.

The basic source of motivation for this article was the incomplete understanding on the role of desorption processes in dark clouds. One of the last modeling studies on the relative significance of several desorption mechanisms was that of \citet{Roberts07}. Since then, several advances in astrochemical modeling have been incorporated in the models on a common basis. They include, first, real time-dependent cloud core models with changing physical parameters \citep{Brown88,Garrod06} and, second, the consideration of subsurface layers of interstellar ice mantles (three-phase models) \citep{Cuppen07}. Additionally, reliable information from experiments or theoretical studies has been published on desorption mechanisms, such as reactive desorption \citep{Garrod07} and photodesorption \citep[e.g.][]{Oberg09a}.

These new data have been implemented into recent astrochemical models \citep[e.g.]{Garrod11,Garrod13a,Du12,Vasyunin13a,Vasyunin13b,Chang14}. Reactive desorption has been analyzed in detail as a non-selective mechanism \citep{Garrod06a,Garrod07,Vasyunin13b}, only, although other papers describe it as a selective mechanism \citep{Du12,Reboussin14}. All these authors have used a two-phase gas-grain model. In the case of desorption by interstellar and cosmic-ray-induced photons, a standard yield of 0.001 desorbed molecules per photon is typically used in the abovementioned papers. Instead, values closer to 0.002 or 0.003 are supported by experiments (see section~\ref{r-pdes}). Finally, desorption by the exothermic H+H reactions on grains was identified as probably the most important desorption mechanism in dark cores by \citet{Roberts07}, yet, it has not been included into any of the recently published models altogether.

Following the above discussion, the aims of this paper are to analyze (1) the significance and selectivity of higher-yield photodesorption (section~\ref{r-pdes}); (2) the effects of selective reactive desorption (section~\ref{r-rdes}); and (3) the importance of indirect reactive desorption, manifested by the H+H reaction (section~\ref{r-hfdes}). The analysis has been performed with a time-dependent three-phase model. This allows to offer a fresh view for aims (2) and (3), and is a requirement for an up-to-date investigation for (1).

\subsection{Ice physical model development}

In order to perform an up-to-date analysis of the desorption mechanisms, an advanced astrochemical model is required. The models employed for the physical description of interstellar ices have often been rather limited. Ices accumulate in mantles with a thickness of around $10^2$ monolayers (ML). The (inert) bulk of the icy mantles was first considered by \citet{Hasegawa93a}. Yet many studies continue to rely on two-phase (gas and surface) models. Any reactions in ice mantles have to be hampered by the obvious difficulties of molecule movement in ice lattice. \citet{Kalvans10,Kalvans13a} have investigated the possible result of mantle reactions that occur on the surface of cavities within the mantle. The cavities contain up to a few per cent of ice molecules at a given time, which means that mantle reaction efficiency is at most, only tenths of that for surface reactions. \citet{Garrod13a} employs a different approach -- mantle molecules are more strongly bound into the ice, and thus move more slowly, with the reaction rates affected accordingly. These are, literally, three-phase models, i.e., the sub-surface mantle was considered as a fully mixed phase with no distinction between ice monolayers. This is a major approximation because it has been shown that ice composition is depth-dependent \citep{Garrod11}.

An effective option for ice-mantle modeling is the Monte Carlo technique. This approach initially suffered from a limited set of surface species and reactions, owing to the computationally intense calculations \citep{Cuppen07,Cuppen09}. The computational cost places some restrictions for more recent models, too \citep{Vasyunin13a}. The approach has been used to validate the results of the macroscopic rate-equation models. One of the main conclusions is that rate-equation models may successfully reproduce chemistry on interstellar grains, only, if molecule binding energies $E_b$ are sufficiently high, e.g. $E_b \geq 0.5E_D$, where $E_D$ is the adsorption (or desorption) energy of the particular species. High $E_b$ means that for reactive species accretion from the gas is faster than the time to scan the surface of the whole grain \citep{Vasyunin09}.  A study of Monte-Carlo random-walk technique application for bulk-ice chemistry is has been published by \citet{Chang14}.

In addition to the investigation of desorption mechanisms, a secondary aim is to develop further the ice-mantle description with the addition of a depth dimension for the subsurface layers. An approach similar to that presented by \citet{Garrod13a} was used. The rate-coefficient calculation for mantle reactions is revisited because the approach of \citet{Garrod13a} has been taken directly from the description of surface chemistry. For surface, an underlying assumption is that molecules can freely migrate. Instead, for mantle chemistry the assumption used here is that molecules \textit{cannot move}, they are mostly frozen in place.

\subsection{Significance of desorption of ice molecules}
\label{i-sig}

Many surface chemistry studies have been performed without a proper reproduction of the observed abundances for important ice species. In particular, the production of CO$_2$ has been a major problem \citep{Ruffle01b} that has recently been solved by \citet{Garrod11}. The `appearance' of CO$_2$ ice at a threshold $A_V$ value was linked to the molecule binding energy, while residing on surface. The synthesis of CO$_2$ is also certainly connected to the rate of accumulation of oxygen and carbon in ice, which is largely governed by non-thermal desorption mechanisms \citep{Leger83}.

Ice composition as a function of desorption has been repeatedly re-visited \citep[e.g.,][]{Leger85,Willacy93,Roberts07}. Several major advances in the field require a renewed dedicated investigation. These are the recognition of the role of the interstellar UV radiation field (ISRF) in regulating ice accretion \citep{Garrod11}, the possibility of subsurface molecule photodesorption \citep{Andersson08,Oberg09b}, codesorption \citep{Oberg09b}, and the possible importance of reactive desorption \citep{Du12,Vasyunin13b}. 

Major interstellar ice components have been detected at different ISRF extinctions. The respective threshold $A_V$ ($A_{\rm th}$) for H$_2$O, CO$_2$, and CO ices are approximately 3.2, 4.3, and 6.8mag, respectively \citep{Whittet01,Whittet07,Bergin05}. These values show a dependence on the desorption energies $E_D$ for these species. Photoprocess, the principal desorption mechanism at low $A_V$, usually is assumed to be non-selective. This is caused by the non-thermal nature for this desorption mechanism and codesorption of different species in an ice mixture \citep{Bertin13}. This discrepancy hampers the attempts to have a full agreement between models and observations \citep{Garrod11,Vasyunin13a}.

Thus, a major aim of this paper is the investigation of reactive desorption and photodesorption that are $A_V$-dependent, and likely govern the accumulation of ice layers. These mechanisms have previously often been assumed being non-selective. The study involves empiric or semi-empiric equations for a proper calculation of desorption rate coefficients.

\subsection{Desorption mechanisms}
\label{i-des}

In the following paragraph, I shortly review desorption mechanisms, used in the present study.

\emph{Evaporation} from grains, which are in thermal equilibrium with the gas, is known to be an effective desorption path for molecules with very small $E_D$ (H, H$_2$, He), or at higher temperatures. It is generally not sufficient for explaining gas-phase molecular abundances in dark clouds, which require other desorption mechanisms \citep{Leger83}. After a protostar has been formed in a collapsing cloud core, gas and dust temperatures grow, and evaporation is among the dominant mechanisms \citep{Brown88}.

\emph{Desorption by the ISRF photons} has often been omitted in models of dark cloud cores because the radiation field is very low at high extinctions ($A_V \geq 10$mag). However, the freeze-out stage for volatile species onto grains is largely governed by the ISRF. As the dense cores evolve, the ice-forming molecules either disappear into the central body or are sublimated by the energy released by the newborn protostar. These processes have been fully reflected in astrochemical models only relatively recently \citep[e.g.,][]{Garrod06}. One of the results is that the lifetimes of the ice mantles are largely determined by interstellar-photon-induced desorption.

Cosmic ray particles that interact with the abundant molecular hydrogen generate an internal UV radiation field within the clouds \citep{Prasad83}. The resulting \emph{secondary photons} are an important internal ionization source, and are able to induce desorption of ice molecules from grains \citep{Duley89,Hartquist90}.

Desorption by photons has traditionally been treated as a non-selective mechanism with uniform yields for all species \citep[e.g.,][]{Roberts07}. The use of molecule-specific yields has become possible with the publication of reliable experimental results \citep{Oberg09a,Oberg09b,Bahr12,Fayolle13}. However, the phenomenon of codesorption may mean that this process is indeed non-selective, at least for species with similar physical properties \citep{Bertin13}. Wavelength-dependent studies \citep{Fayolle13,Bertin13} have revealed that the yield is different for conditions at cloud edges and cloud center, where ISRF and cosmic-ray-induced photons dominate, respectively.

%Because the energy delivered to grain by a single hit is high, this mechanism -- as local heating -- may work even for large and coagulated grains \citep*{Leger85}.
\emph{Whole-grain heating} by cosmic-ray iron nuclei and subsequent ice sublimation has been recognized as one of the principal desorption mechanisms in dark cloud cores \citep*{Leger85}. These authors consider evaporation of CO from grains heated above 25K. In another classical paper by \citet*{Hasegawa93a}, the evaporation is approximated as occurring on grains at $T_{\rm warm}=70$K for a time $t_{\rm warm}=10^{-5}$s after each hit of iron cosmic-ray nuclei. The latter approach has often been used by subsequent astrochemical models.
%\emph{Desorption from localized spots} on grain surface, created by a passing cosmic-ray particle, can be described as evaporation \citep{Watson72a,deJong73,Leger85}, sputtering of ice molecules \citep{Draine79}, or ejection of ice fragments \citep{Johnson91,Duley96}. Of these, only evaporation is a selective process.
%Local heating affects a certain region within the grain -- a cylinder along the ion's path. It lasts for a very short time ($\approx10^{-10}$s), before the heat is dissipated to other parts of the grain. This process has a low efficiency, with desorption yields usually not exceeding $10^4$ molecules per ion. Most of the heat is lost via the whole-grain heating process \citep[for grains with radius $a<0.2\mu$m, ][]{Leger85}. Consequently, local heating usually is not applied for chemical models of dark cloud cores.
%\emph{X-rays} cause effects, similar to those of fast ions, and may desorb even species with high desorption energies, such as methanol or water \citep{Andrade10,Meijerink12,Mendoza13}. Because the molecules are almost exclusively desorbed in the form of radicals and ions, it is reasonable to classify X-ray photons as a non-selective mechanism. X-rays are likely to be subjected to attenuation by gas and dust, and are usually not considered in dark cloud core models. Protostars may emit large amounts of X-ray photons, which dominate the ionization balance over cosmic rays, and their effects have to be considered in the case of protoplanetary disks \citep{Casanova95,Gorti09,Andrade10}.

\emph{Indirect reactive desorption} by the excess energy of exothermic reactions on grains has been found to be a potentially important mechanism \citep{Williams68,Watson72a,Watson72b}. It is driven by ionizing radiation fields that generate radicals and ions in gas and ice. The subsequent surface reactions release the energy, initially delivered by photons or cosmic rays. This type of desorption is tied to the radical content in the cloud, which is determined by ionizing radiation flux in it. Thus, reactive desorption is an $A_V$-dependent process up to a certain depth, when cosmic-ray induced effects start to dominate. This mechanism can be very important regarding the formation of ices \citep{Garrod11}.

As a separate desorption mechanism, the \emph{formation of molecular hydrogen} on grains was considered by \citet{Duley93} and \citet{Willacy94}. In addition to the ejection of the newly formed H$_2$ molecule, excess energy transferred to ice lattice may result in the desorption of a nearby heavy atom or molecule. The excess energy is of order a few eV \citep{Roberts07}, much less than that of UV photons. The yield is currently unknown, and is more constrained by its effect on modeled molecular abundances in clouds. This process has been assumed to be effective only for species with low $E_D$, e.g., for CO and lighter species, and thus it is a selective desorption mechanism.

%\emph{Chemical explosions} have been proposed as a potential desorption mechanism for interstellar ice molecules \citep{Greenberg73}. Experiments show that UV-irradiated interstellar ice analogues explode when their temperature is allowed to increase. This is because the irradiation causes an accumulation of radicals in ice lattice. When a major part of the ice is heated to $\geq$27K they become sufficiently mobile to produce an exothermic radical chain reaction \citep{dHendecourt82}. The resulting explosion effectively releases the whole ice layer, or large parts of it, into the gas phase. It happens likely via the ejection of whole clusters of ice molecules. The radical content in the cold ice reaches a few per cent at most \citep{Schutte91}. The heating necessary for the initiation of the chain reaction can be provided by grain collisions \citep{dHendecourt82} or energy  absorption from a X-ray photon, or a cosmic-ray particle \citep{Leger85}.
%Because the explosions affect the whole ice mantle at once, they can be regarded as a non-selective desorption mechanism. They can be important for the production and release of heavy species, e.g., complex organic molecules \citep{Sorrell97,Cecchi10,Rawlings13}. This mechanism has often been omitted in less-specific astrochemical modeling papers, and has not been considered in the current paper.

Other possible desorption causes include molecule ejection locally by cosmic rays, X-rays (important in protostellar disks), and chemical explosions. Of all these mechanisms, reactive desorption, desorption by the H+H reaction and whole-grain heating are truly selective. The latter is rather ineffective. Experiments have shown that photodesorption (ISRF and cosmic-ray-induced) displays some selectivity, too \citep{Oberg09b,Fayolle13,Bertin13}. Thus, direct and indirect reactive desorption in addition to photodesorption can be considered as selective and important during the formation of interstellar ices. These processes are investigated further in this paper.

In section~\ref{mod} a description of the chemical and physical model is provided. Each desorption process has been studied individually in Sections~\ref{r-pdes}, \ref{r-rdes}, and \ref{r-hfdes}, respectively. In section~\ref{r-compl} the results of a model that combines elements of previous simulations are investigated. Concluding remarks are summarized in section~\ref{concl}.

\section{The model}
\label{mod}

\subsection{Sublayer approach in ice modeling}
\label{sublay}

A modified code of the ALCHEMIC astrochemistry program \citep{Semenov10} was utilized to perform chemical kinetics calculations. Ice-related processes have been added (e.g., desorption mechanisms, the formation of subsurface ice mantles, and molecule diffusion in ice), and a real time-dependency, with changing temperature, density, $A_V$, and grain size, has been introduced into the model.

An important feature of the model is the treatment of subsurface ice mantle. It has been recently recognized that molecules in ice lattice may change their location as a result of thermal diffusion \citep{Oberg09c,Fayolle11a}. Meanwhile, reactions are taking place directly in ice lattice, too, evidenced by the formation of different molecules in ices in the presence of radicals or other reactive species \citep[e.g.,][]{Gerakines96,Oberg11a,Linnartz11,Noble13}. This has been reflected in a recent astrochemical model by \citep{Garrod13a} that includes reactions among molecules in subsurface ice. More rigorous methods are those based on the Monte-Carlo technique, although they may have a high computational cost \citep[e.g.,][]{Vasyunin13a,Chang14}.

In order to include subsurface ice chemistry into the simulation, ice was described as consisting of four layers -- the surface and three mantle layers. Such an approach lacks the resolution of \citet{Hasegawa93b} and similar models that consider ice monolayers. Calculating chemical reactions in each monolayer for each time-step of the simulation is and enormous computational task. This is probably unnecessary, given that interstellar ices likely are characterized by irregularity.

Ice accumulates onto irregular grains, it is amorphous, and may contain inhomogeneities, such as pores, large organic species, PAHs, or even small grains. Photodissociation products almost instantly migrate across several monolayers \citep{Andersson06,Andersson08}. A similar effect is possible for the products of exothermic binary reactions. In such an irregular ice, it is hard to define the boundaries and properties of a monolayer. Additionally, a radical in ice may react with any of its neighbors, some of which are in different monolayers, and it is unclear, which monolayer should the product species belong to. Most of these irregularities are of no concern, if ice layers of ten or more monolayers thick are considered as whole entities.

Thus, it can be physically adequate to employ ice model that describes the ice as consisting of several subsurface layers instead of hundreds of monolayers. For short, I refer to these former layers as `sublayers' hereafter. Each sublayer  may contain a few up to dozens of monolayers. Molecules are assumed to be intimately mixed within a sublayer. Chemical reactions are permitted between same-sublayer species, only. Molecule exchange via diffusion can occur between adjacent sublayers.

The number of sublayers can be chosen depending on the aims of the work. A minimum number of two is necessary to represent the two most distinct ice environments -- polar H$_2$O-dominated and non-polar CO-dominated ices \citep{Sandford88,Tielens91}. Three or more sublayers would be more adequate to perform calculations with the resolution required for a model that is able to fully account for ice observations in dense cores. This is much less than the hundreds of monolayers that have to be considered for some of the other bulk ice models \citep{Garrod11,Taquet12,Vasyunin13a}. The low number of sublayers allows the explicit consideration of subsurface reactions in a depth-dependent manner. Three was chosen as the number of subsurface layers for the current research. They were numbered, beginning from the surface, with sublayer 1 being the outermost and sublayer 3 the innermost layer (connected to the grain itself). The outer surface is a separate, `zeroth', layer. The formation of ice is described in section~\ref{mantevo}.

\subsection{Physical model}
\label{phys}

The rates for a physical phase-change or ice layer transition processes and chemical reactions are calculated via classical chemical kinetics equations. The rate coefficients for specific processes are given in sections below. In cases, when phase-change occurs, we use the denominations $f0$ and $f$ to indicate the initial and final phases, respectively. In this regard, the `phases' are gas ($g$), surface ($S$), mantle sublayer 1 ($M1$), sublayer 2 ($M2$), or sublayer 3 ($M3$).

\subsubsection{Cloud conditions}

The model considers a gas clump in a cloud core in isothermal collapse. This approach has been adopted from \citet{Garrod06}. The temperature $T$, dependent on $A_V$, was calculated according to \citet{Garrod11}. The time-dependent results may also be interpreted as being dependent on the distance from the center of the contracting core. In such a case, earlier times represent gas clumps that reside further from the center, similarly to the model by \citet{Garrod08}.

The physical conditions were adopted for a 2$M_\odot$ core and initial density $n_{\rm H}=3\times10^3$cm$^{-3}$. The density increases to 10$^7$cm$^{-3}$ over a period of 1Myr, in line with \citet{Brown88} and \citet{Nejad90}. It was assumed that a screen of diffuse gas with $A_V=1$ surrounds the contracting core. The visual extinction to the center of the clump was calculated self-consistently, and grows from below 2 to over 200mag. The physical conditions in the cloud are similar to those figured in \citet{Vasyunin13a}. The self- and mutual-shielding of CO and H$_2$ has been included with the use of the tabulated data from \citet{Lee96}. The same was done for the N$_2$ molecule with the use of tabulated data from \citet{Li13}.

The abundances of the chemical elements have been adopted from \citet{Garrod06}. Because a partially shielded molecular core is modeled, all elements, except for hydrogen, initially are in neutral atomic form in the gas phase. Hydrogen is divided between molecular H$_2$ (99.9\%) and atomic H (0.1\%).

Hydrogen ionization rate by cosmic rays was assumed characteristic for dark clouds, $1.3\times10^{-17}$s$^{-1}$ \citep{Tomasko68}.  The flux of cosmic-ray induced photons was taken to be $F_{\rm crph}=4875$s$^{-1}$cm$^{-2}$, and the flux of heavy cosmic rays $F_{\rm FeCR}=2.06\times10^{-3}$ s$^{-1}$cm$^{-2}$ \citep{Roberts07}. Finally, the assumed flux of ISRF far-ultraviolet photons for photodesorption calculations was $F_{\rm ISRF}=1.7\times10^{8}$s$^{-1}$cm$^{-2}$ at $A_V=0$mag \citep{Tielens05}.

\subsubsection{Grain properties and grain-related effects}
\label{grs}

The 'standard' reaction network of \citet{Laas11} was used for gas-phase reactions, gas-grain interactions and solid-phase reactions. This database, in turn, was based on the work by \citet*{Garrod08}. Three changes were introduced -- H$_2$ adsorption energy was taken from \citet{Katz99}, energy barrier for the CO~+~O surface reaction was taken 290K from \citet{Roser01}, and the gas-phase reaction CH$_3$O + CH$_3$ was added from \citet{Vasyunin13b}. Following the experimental results of \citet{Katz99}, quantum tunneling for diffusion and reactions was not considered. For reactions and physical processes, the model considers grains with a radius $a=0.1\mu$m and an additional ice thickness $b$. The abundance of grains $n_g=1.3\times10^{-12}n_{\rm H}$ was calculated for spherical particles with a density of 3g~cm$^{-3}$ that contain 1\% of cloud mass. It was assumed that the grains are rough, having $N_s=1.5\times10^6$ adsorption sites on the surface. This number was kept constant during cloud evolution under the assumption that surface roughness is smoothed out by the accumulating ice layers, while the size of the grain increases. Where necessary, it was assumed that a `cubic average' ice molecule has a size of $b_m=3.7\times10^{-8}$cm. Maximum ice thickness in monolayers $B=b/b_m$, achieved by this model, does not exceed 180.

\subsection{Gas and gas-grain interactions}
\label{gasgr}

Gas-phase processes include binary reactions, dissociation and ionization by interstellar and secondary photons, and ionization by cosmic rays. The grain albedo was taken to be 0.5. Gas-grain interactions consist of (dissociative, non-adsorptive) recombination between positive ions and negatively charged grains, molecule accretion (freeze-out), and desorption. These processes were included as in the ALCHEMIC model \citep{Semenov10} for grains with a radius $a+b$, with $b$ calculated self-consistently. The sticking coefficient was taken unity for heavy species and 0.33 for hydrogen atoms and molecules \citep{Brown90}.

In the reference, or `Standard', model four desorption mechanisms, ($f=g$, $f0=S$), have been included. They are evaporation at the equilibrium grain temperature $T$, evaporation at 70K, induced by whole-grain heating, \citep{Leger85,Hasegawa93a}, photodesorption by interstellar photons \citep{Turner98}, and by cosmic-ray induced photons \citep{Roberts07}. Additionally, reactive desorption was considered, assuming that 1.0\% of surface reaction products are released into the gas \citep{Garrod07}. These mechanisms for the Standard model have been chosen to be in line with recent models by other authors \citep{Garrod11,Garrod13a,Vasyunin13a,Albertsson13}.

The desorption yield for the secondary photons has been measured or calculated for several species \citep{Andersson08,Oberg09a,Oberg09b}, while the yield for most other species is unknown. In order to avoid unnecessary bias, an uniform yield of $Y_{\rm crph}=10^{-3}$ desorbed molecules per photon was assumed for all species in the Standard model. There are experimental results that indicate a higher desorption yield for higher-energy photons \citep{Bahr12,Fayolle13}, although it is dependent on the absorption spectra of a particular species \citep{Bertin13}. For consistency, it was assumed that $Y_{\rm crph}=Y_{\rm isrf}$. The rate coefficient for photodesorption is
   \begin{equation}
   \label{gas1}
k_{\mathrm{pd}}=\frac{\pi(a+b)^2 F_{\mathrm{ph}} Y_{\mathrm{ph}}}{N_s},
   \end{equation}
where `ph' denotes either ISRF or cosmic-ray-induced secondary photons and $a+b$ is the time-dependent grain radius (section~\ref{grs}). The fixed value of $N_s$ means that photodesorption efficiency grows with the number of adsorbed molecules (up to 3ML, see below). This takes into account the localized nature of photon absorption in a grain \citep[e.g.,][]{Duley88} and the dependence of photodesorption yields on ice depth \citep{Andersson06,Andersson08,Oberg09b,Bertin13} in a simple and natural way within the model.

Photodesorption has been attributed to all sublayer species, given that the number of above monolayers does not exceed two. The desorption rate from a sublayer was divided by the number of monolayers in that sublayer. This accounts for the fact that only the upper ML has been exposed to desorption. Thus, desorption from a depth of up to three MLs is possible. This approach means that a number of ice monolayers can accumulate onto grains at higher $A_V$ values, earlier in the evolution of the cloud core, before photodesorption reaches its maximum effectiveness.

In addition, desorption from the molecular hydrogen formation reaction on grains was included in the model for case investigation (section~\ref{r-hfdes}). The selectivity of ISRF, secondary-photon induced photodesorption, and reactive desorption was investigated in section~\ref{res}, too.

\subsection{Physical description of ice}
\label{mantevo}

\subsubsection{Properties of ice layers}
\label{iceprop}

It was assumed that ice consists of a total of four layers -- the surface (abbreviated $S$), and three subsurface mantle layers, or sublayers 1, 2, and 3 ($M1$, $M2$, and $M3$, respectively). $M1$ lies beneath the surface, while $M3$ is bound to the grain core. Molecules within each of the layers are completely mixed together. This approach is in line with recent similar models that consider subsurface ice chemistry with the rate-equation method \citep{Garrod13a,Kalvans13a}. These authors describe models with a single mantle layer (one sublayer), only. This causes the model to be inaccurate. A major problem for such a model is, e.g., the interaction of CO, supposed to be in the outer layers, with the photodissociation products of H$_2$O in the inner layers. The inclusion of several sublayers removes this and similar discrepancies, and can be considered as a novelty. Additionally, the sublayer approach allows to perform a limited analysis on the depth-dependent composition of the ice layer (section~\ref{r-sub}).

The rate of chemical reactions and species' interchange between the layers is largely governed by molecular diffusion. The diffusion rate is dependent on the binding energy $E_b$ of each molecule. In the case of surface species, the approach by \citet{Garrod06} was used, i.e., $E_{b,S}=0.5E_D$.

For subsurface species, several differences have to be taken into account. First, unlike surface species, molecules in the bulk are surrounded by other molecules, and cannot directly desorb into the gas. This means that the term `desorption' energy is fictional; in fact, it is the \textit{absorption} energy $E_B$. By definition, $E_D$ or $E_B$ is the energy required to break all physical bonds for a molecule on ice, or within ice lattice. It is used to calculate the species' vibrational frequency, $\nu_0$, Eq.~(\ref{evo2}).

\citet{Garrod13a} assumed that $E_D$ or $E_B$ are equal. However, the number of neighboring species in the lattice is higher than that for surface species, and $E_B$ can be expected to be correspondingly higher than $E_D$. The precise $E_D/E_B$ ratio probably should depend on the average number of species surrounding a molecule on surface and in bulk ice. This depends on small-scale surface roughness and ice porosity. Additionally, the molecular composition of ice is important, because strongly polar species (e.g., H$_2$O, HCOOH, NH$_3$) are able to form much stronger intermolecular bonds than molecules with a low polarity (e.g., CO, CO$_2$, N$_2$).

A constant each species' $E_D$ for the surface layer and $E_B$ for each of the sublayers was adopted, in line with previous studies \citep[e.g.][]{Hasegawa92}. It was assumed that $E_{B,M1}=3.0E_D$ and $E_{B,M2}=E_{B,M3}=3.3E_D$ for sublayers 1, 2, and 3, respectively. $E_B$ for the inner sublayers has been assumed higher than that of sublayer 1, because a major part of these sublayers consist of the polar H$_2$O molecules, while sublayer 1 has a high proportion of CO after the freeze-out process has ended. In the present study, the difference between $E_{B,M1}$ and $E_{B,M2}$, $E_{B,M3}$ was taken to be mediocre 10\%. Such a cautious approach was chosen to ensure that chemical reactions are possible in all ice layers, corresponding to photoprocessing experiments with astrophysical ice analogs \citep[e.g.][]{Gerakines96}. For mantle binding energy $E_{b,M}$ the approach of \citet{Garrod06} was retained, i.e., $E_{b,M}=0.5E_B$. The high $E_B$ values imply a very slow diffusion, the molecules are basically frozen in place at temperatures relevant for this model (section~\ref{mandiff}).

In addition to the above discussion, binding energy of mantle species must be higher than desorption energy of surface species, i.e. $E_{b,M}>E_D$. This is to ensure that the movement of bulk ice species is slower than evaporation, a necessary condition for the ice to be described as a solid. This condition was not recognized by \citet{Garrod13a}. In a case with $E_{b,M}<E_D$, ice molecules become mobile before evaporation, forming a liquid, which is not physically justifiable for interstellar or circumstellar conditions.

The desorption energy for surface species can be dependent on the fraction of the surface consisting of H$_2$ molecules. Species, adjacent to a H$_2$-covered surface may have much lower actual $E_D$ than species attached directly to ice surface. In the present study, this was attributed to the light H and H$_2$ species, only, with the approach proposed by \citet{Garrod13a}. It was not attributed to heavy-atom containing species. In other words, heavy molecules do not 'step on to' H$_2$ when accretion or diffusion occurs. Instead, the mobile hydrogen species make room and allow the heavy molecule to be always firmly bound to the ice surface. The vibration frequency of a heavy species, adjacent to H$_2$, is orders of magnitude higher than the diffusion rate of H$_2$. When the adjacent H$_2$ molecule hops away, no other H$_2$ molecules will be able to move into the now free adsorption site before the adjacent heavy molecule approaches the site via vibration, and binds itself strongly to the surface.

\subsubsection{The formation of the ice layer}
\label{comp}
%Figure 1
\begin{figure*}
% \vspace{4cm}
  \hspace{-1cm}
  \includegraphics[width=18.0cm]{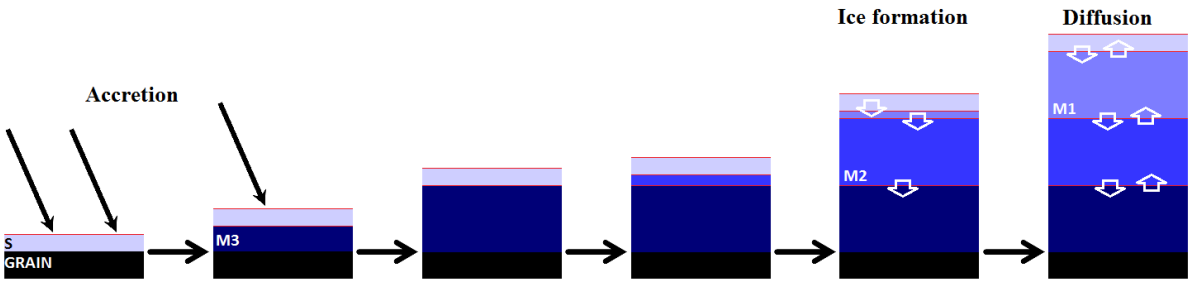}
 \vspace{-20cm}
 \caption{Schematic drawing of the ice mantle on an interstellar grain as considered in the model. The allowed molecule transfers between the four ice layers are shown. In ice formation (`compaction') process, molecules are sequentially transferred from the gas to the surface $S$ and, then, the sublayers $M1$, $M2$, and $M3$, as long as net accretion happens. The molecules start to `pile up' in the inner sublayers, up to a thickness of 60ML. The bi-directional diffusion of ice species is permitted at all times between adjacent ice layers. (A color version of this figure is available in the online journal.)}
 \label{att-bilde}
\end{figure*}
Figure~\ref{att-bilde} shows the ice layer, as it is described in the model. The formation of ice starts with the accretion of species onto grains. This forms the surface layer. If the thickness of the surface layer exceeds 2ML, the molecules are transferred to sublayer 1. They are then transferred to sublayer 2 and, after that, to sublayer 3, if their thickness does not exceed 60ML. This means that, initially, the molecules are transferred through sublayers 1 and 2, and pile up in sublayer 3. After the abundance of molecules in $M3$ has reached the equivalent of 60ML, sublayer 2 is filled up to the same thickness. All the sublayers reach an approximately similar thickness after the accretion period has ended (final ice thickness 160-170ML, depending on the particular desorption model).

The transition between the surface layer and sublayer 1 is of particular concern, because the chemical processes differ greatly in these two phases. Molecules in the surface layer are able to freely diffuse around, while they are mostly frozen in place in the case of mantle layers (section~\ref{icech}). Diffusion experiments with interstellar ice analogs \citep{Palumbo06,Raut07b} have shown that molecules are highly mobile in porous ice layers, i.e., the surface area of ice is higher than that of a single smooth monolayer on top of the ice. This means that the nominal thickness of the surface layer is not limited to a single monolayer.

\subsubsection{Ice compaction: transition of species from surface to mantle}

Experiments \citep{Palumbo06,Raut07a,Raut07b,Accolla11} have shown that interstellar ice analogs are porous, when obtained by depositing molecules onto cold surfaces. Ices obtained by deposition of \textit{free atoms} and subsequent reactions on the surface, have a compact structure \citep{Oba09}. Additionally, it has been demonstrated by detailed models that slowly accumulating ices have a compact structure, while rapidly accreted ice can be a highly porous \citep{Garrod13b}.

Based on these evidences, it is possible to make out two `modes' for interstellar ice structure. The first is water-dominated ice, which forms by the slow accretion of H, O and other atoms, followed by reactions on the surface. This mode dominates during the long low-density period ($n_{\rm H}\leq2\times10^{4}$cm$^{-3}$ in core contraction. The second mode sets in as the contraction is accelerated and gas density exceeds $10^{4}$cm$^{-3}$. The majority of species being accreted are CO and N$_2$ molecules that simply accumulate onto the grains with no chemical reactions synthesizing major ice (surface) components. It can be seen that the first mode corresponds to compact ice, and the second mode -- to porous ice layer. The the formation of the subsurface ice has been described in the model with the intention to describe both modes, and the transition between them.

It has been demonstrated that interstellar ice analogs undergo compaction (reduction of porosity), when exposed to UV photons \citep{Palumbo10}, cosmic rays \citep[fast ions][]{Palumbo06,Raut07a}, or exothermic reactions \citep{Accolla11}. Additionally, observational evidence suggests that actual interstellar ices are compact in nature \citep{Palumbo10}. Because of this, the nominal thickness of the surface layer (representing porous ice) was never allowed to become thicker than 10ML. In the model, the transition of species from the surface layer to the compact sublayer 1 was described as a first-order process. The rate coefficient was found empirically:
   \begin{equation}
   \label{phys1}
k_{\mathrm{comp}}=5\times10^{-13}B_S^2, s^{-1} ,
   \end{equation}
where $B_S$ is the thickness of the surface layer in MLs. The rate coefficients for other transitions associated with ice formation in the sequence $S \rightarrow M1\rightarrow M2\rightarrow M3$ were calculated with Eq.~(\ref{phys1}), too.

The compaction process is initiated only, when surface ice thickness exceeds 2ML. This is consistent with the finding by \citet{Vasyunin13a} that more than one reactive surface monolayer is necessary to properly describe surface chemistry. Based on pre-calculated ice maximum thickness, $B<180$ML, it was assumed that each sublayer has a maximum thickness of 60ML. Because the inner layers are filled first, sublayer 1 does not get full in the models described here. The result of the mantle formation process can be seen in figures \ref{att-sub} and \ref{att-r-subab}.

\subsubsection{Hydrogen diffusion}

Most of ice species diffuse between the ice layers $S, M1, M2$, and $M3$ via thermal hopping (section~\ref{mandiff}). The diffusion of hydrogen species H$_2$ and H was treated separately -- with the use of data derived from experiments and dedicated models. This approach is in line with the earlier studies by \citet{Kalvans10,Kalvans13a}.

The hydrogen diffusion coefficient in ice $D$ (cm$^2$s$^{-1}$) can be expressed as
   \begin{equation}
   \label{arrh}
D = D_0 e^{E_{\rm act}/T},
   \end{equation}
where $E_{\rm act}$ (K) is the activation energy for diffusion \citep{Strauss94}. The parameters $D_0$ and $E_{\rm act}$ for H$_2$ were calculated from data supplied by \citet{Strauss94} and \citet{Awad05}, as specified in \citet{Kalvans13b}. The adopted parameters are $D_{0,H}=10^{-12}$, $D_{0,H_2}=4.76\times10^{-4}$cm$^2$s$^{-1}$, $E_{\mathrm{act},H}=185$, and $E_{\mathrm{act},H_2}=90$K.

The temperature-dependent value of $D$ is then used to calculate the diffusion rate coefficient, according to Eq.(19) of \citet{Kalvans13a}. The time-dependent thickness of the sublayers and the surface layer was taken into account for this calculation. The permitted transitions between ice layers for hydrogen diffusion are $S \longleftrightarrow M1 \longleftrightarrow M2 \longleftrightarrow M3$.

Because the diffusion of H$_2$ is relatively fast, the maximum proportion of H$_2$ in ice has to be regulated. It can be reasonably assumed that most of the hydrogen resides in cavities within the ice \citep{Strauss94}. The estimated porosity of interstellar ices is around 5\% \citep{Oberg11b}. Following this, it was assumed that the diffusion of H$_2$ towards a sublayer stops, if the proportion of H$_2$ in that sublayer exceeds 5\%.

\subsection{Ice chemistry}
\label{icech}

\subsubsection{Surface binary reactions}
\label{surf}

Reactions on grain and ice surfaces have been considered with the approach of \citet{Hasegawa92}. The reaction rate was adjusted by reaction-diffusion competition \citep{Garrod11} and modified rate-equations method \citep{Caselli98}. The rate coefficient for binary surface reactions (cm$^3$s$^{-1}$), is
   \begin{equation}
   \label{surf1}
k_{ij}= U_{\mathrm{act}}(ij) (R_{\mathrm{diff},i} + R_{\mathrm{diff},j}),
   \end{equation}
where is $U_{\mathrm{act}}$ the probability for the reaction to occur, related to the activation energy barrier $E_A$. The rate, ($s^{-1}$) at which molecule $i$ scans the whole grain via thermal hopping
   \begin{equation}
   \label{surf2}
R_{\mathrm{diff},i}=k_{\mathrm{hop},i}/N_s.
   \end{equation}
where
   \begin{equation}
   \label{surf3}
k_{\mathrm{hop},i} = \nu_{0,i} \mathrm{exp}(-E_{b,i}/T).
   \end{equation}
is the rate (s$^{-1}$) with which molecule $i$ migrates from one surface site to an adjacent one.

Following \citet{Garrod11}, in the case of reaction-diffusion competition, the reaction probability is
   \begin{equation}
   \label{surf4}
U_{\mathrm{act}}(ij)= \frac{\nu \kappa(ij)}{\nu \kappa(ij) + k_{\mathrm{hop},i} + k_{\mathrm{hop},j}},
   \end{equation}
where $\nu=\nu_{0,i}+\nu_{0,j}$ and $\kappa(ij)=\mathrm{exp}(-E_A/T)$. Reactive desorption \citep{Garrod07} is applied by assuming that one per cent of all reactions products transit to the gas phase. The modified rate-equation method has been applied to all surface reactions, as implemented in the `ALCHEMIC' code \citep{Semenov10}.

The synthesis of CO$_2$ on very cold ($<$8K) grains via O+H$\rightarrow$OH reaction on top of a CO molecule \citep{Garrod11} has not been included in the model. It is a specific process, which has not been attributed to other species and thus may introduce bias in calculation results. Additionally, as recognized by the same authors, molecules on surface can be adjacent to several neighbors. The O+H+CO mechanism requires that an O atom and, half-way into the reaction, an OH radical has attached itself exclusively to a single CO molecule. Other common interactions for the newly formed and still excited radical could include reaction with H$_2$, which is almost omnipresent on cold grains, strong hydrogen-bonding or reaction with the abundant H$_2$O, or even a reaction with CO$_2$. The O+H+CO reaction has been verified as effective by the microscopic Monte-Carlo modeling technique \citep{Chang12}, but, again, their model by design implies that a surface molecule is bound to a single species below it. If OH and CO fail to react immediately at 8K, this may well mean that OH becomes fixed in a configuration in which it is unable to reach CO before it is transformed into H$_2$O by H atoms. H has a non-negligible abundance even in dense cloud cores \citep{Goldsmith05}. Based on these considerations, I suggest that it is acceptable not to use the tertiary reaction for CO$_2$ synthesis in the current model.

\subsubsection{Chemical reactions in sublayers}
\label{manreac}

Due to the relatively high $E_{b,M}$ values ($\geq1.5E_D$, section~\ref{iceprop}), molecules in bulk ice are practically immobile in the temperature range considered. Even light radicals with $E_D=800$K require $>10^{12}$ years to make a single hop at 16K, the maximum temperature permitted in the model. Because of this, one cannot assume that subsurface molecules and radicals are mobile within their respective sublayer, which is the case of the model by \citet{Garrod13a}. A new approach is proposed here, taking into account that molecules are locked in their absorption sites for long timescales.

Species in the mantle were treated as particles adsorbed to the surface of ice lattice cells. Because of the stronger binding of molecules in bulk ice, the characteristic vibration frequency $\nu_{0,f}$ for molecules in sublayer $f$ is different from that of surface species. Following the formalism of \citet{Hasegawa92}:
   \begin{equation}
   \label{evo2}
\nu_{0,f}=(n_s E_{B,f} / \pi^2 m)^{1/2},
   \end{equation}
where $n_s$ is the density of adsorption sites in a lattice cell, $m$ is the mass of the adsorbed particle, and $f$ is either $M1$ for sublayer~1, $M2$ for sublayer~2, or $M3$ for sublayer~3. Molecules in ice lattice cells vibrate with frequency $\nu_{0,f}$. A chemical reaction may occur, if they approach another molecule in the cell and overcome a certain energy barrier, $E_{\rm prox}$. This `proximity barrier' was assumed to arise from the adsorption force exerted by other neighboring species to the molecule that is approaching its reaction partner. Such a parameter has never appeared in the astronomical literature before. It cannot be chosen arbitrary, because it affects the abundance of chemical radicals in ice. It was assumed that $E_{\mathrm{prox}}=0.1E_B$, which results in radical (mainly OH and NH$_2$) abundances $\leq 0.2$\% of total ice. The factor 0.1, multiplied by species' sorption energy in ice, was used also for calculating lateral bond strength between molecules by \citet{Chang12}.

% The calculation of binary reaction rate coefficients for sublayer species was based on the following considerations. Molecules in ice lattice cells vibrate with their characteristic frequency $\nu_{0,f}$, Eq.(\ref{evo2}). They have a limited number of neighbors (assumed $N_c=10$ in this model) that can be reached for chemical reactions. Otherwise, the molecules are immobile. In order to achieve a sufficient proximity for a reaction, a certain energy barrier, $E_{\rm prox}$ has to be overcome. The barrier is assumed to arise from the adsorption force exerted by other neighbor species to the molecule that is approaching its reaction partner. Such a parameter has never appeared in the astronomical literature.

%Unless $E_{\rm prox}$ is high enough to prohibit most reactions ($\approx0.2E_{B}$ or higher), the particular value of this barrier does not significantly affect ice composition. It strongly affects the abundances of chemical radicals in ice. Very low barriers make mantle reaction rates very high, and hamper the calculations. \citet{Chang12} used a value of 0.1$E_D$ for lateral bonds of surface species. In the model, it was assumed that $E_{\mathrm{prox}}=0.1E_B$, which produces radical (mainly OH and NH$_2$) abundances $\leq 0.2$\% of total ice.
The scanning rate of the full surface of a single lattice cell for a molecule $i$ is
   \begin{equation}
   \label{man1}
R_{c,f,i} = k_{\mathrm{hop},f,i}/N_c = \nu_{0,f,i} \mathrm{exp}(-E_\mathrm{prox}/T)/N_c,
   \end{equation}
where $N_c=10$ is the assumed average number of neighbors available for reactions in each lattice cell. The rate coefficient for a binary reaction (cm$^3$s$^{-1}$), in ice mantle lattice cell is
   \begin{equation}
   \label{man2}
k_{f,ij}= U_{\mathrm{act},f}(ij) (R_{c,f,i} + R_{c,f,j}).
   \end{equation}
The reaction probability $U_{\mathrm{act},f}$ is calculated similarly as in the case of surface reactions, Eq.~(\ref{surf4}). The modified rate equations method was not applied in the case of mantle chemistry because there is no accretion and no desorption in the bulk ice, new species arrive mostly via the photoprocess, and, in terms of this model, the reactive volume consists of a large number of interconnected lattice cells. Such a regime has not been tested by the Monte-Carlo approach.
%A modified rate-equation approach designed for surface reactions is unsuited for direct application for mantle chemistry. The mantle is not considered as a single entity. Thus, the modified rate equations are applied for surface reactions, only.
%Molecule thermal diffusion was considered in the model in reaction-diffusion competition for reactions in the sublayers.

\subsubsection{Diffusion of species in ice mantles}
\label{mandiff}

Molecule thermal diffusion between adjacent sublayers, negligible for temperatures in the present study, can be of high importance for ices in protostellar envelopes \citep{Garrod13a}, and thus was included in the model. The permitted directions for inter-sublayer diffusion are $S \longleftrightarrow M1 \longleftrightarrow M2 \longleftrightarrow M3$. Diffusion rate coefficients are calculated according to
   \begin{equation}
   \label{evo3}
k_{\mathrm{diff}}= \frac{\nu_{0,f0} \mathrm{exp}(-E_{b,f0}/T)}{yB_{f0}},
   \end{equation}
where $B_{f0}$ is the actual thickness (in MLs) of the source layer ($S, M1, M2$, or $M3$), and $y$ is the probability for a molecule to move one ML towards the target layer. In other words, $y$ is the number of steps required for a molecule to move closer  by one ML towards its target layer. Assuming a simple, cubic symmetry we get $y=6$. The exact value of $y$ is insignificant, because bulk diffusion is very slow. The above means that $yB_{f0}$ is the total number of steps for a molecule for the diffusion into an adjacent ice layer.

$E_b$ for molecules in the surface layer is much lower than that of mantle species. Because of this, surface species would diffuse much faster from surface to sublayer~1 than in the opposite direction. This is unrealistic, because the diffusion of molecules between ice layers involves `making room' in the target layer \citep[molecule swapping,][]{Fayolle11a}. Taking this into consideration and to reflect the diffusion process more precisely, I assumed that in the case of surface molecules, $E_{b,f0}=E_{b,M1}$ for equation~(\ref{evo3}). This ensures an similar, slow diffusion rate between for the directions $S \longrightarrow M1$ and $M1 \longrightarrow S$.

%$E_b$ in the surface layer is much lower than that of sublayer 1. In order to avoid excessive diffusion of molecules into sublayer 1 from the surface, the value of $E_{b,f0}$ was taken to be equal to $E_{b,M1}$ for the transition $S \longrightarrow M1$. This is justified by the fact that the diffusion of molecules between the ice layers involves leaving the source layer and `making room' in the target layer \citep[molecule swapping,][]{Fayolle11a}. This means that the rate of molecules hopping from one layer to another depends on the binding energies of molecules in both layers. Any ice layer can become either thicker or thinner as a result of diffusion, although the total effect is negligible because of the high binding energies.

\subsubsection{Dissociation of ice species}
\label{diss}

Dissociation by interstellar and cosmic-ray-induced photons has been included for gas, surface, and sublayers. For species in the icy mantles below the surface, the attenuation of radiation by overlying ice layers has to be considered. Each sublayer has a different (and time-dependent) number of overlying monolayers. For surface molecules, the attenuation factor was assumed unity. For subsurface species, the attenuation factor was calculated for the middle monolayer in the respective sublayer for each time-step. It was then attributed to all species in the particular sublayer. Taking into account that each monolayer has an absorption probability of $P_\mathrm{abs}=0.007$ \citep{Andersson08}, the attenuation factor is
   \begin{equation}
   \label{diss1}
A_\mathrm{ice}= (1-P_\mathrm{abs})^{B_\mathrm{a}+0.5B_f},
   \end{equation}
where $B_{\rm a}$ is the number of monolayers above the sublayer in consideration, and $B_f$ is the number of monolayers in this sublayer.

Dissociation products remain in their parent sublayer. It was assumed that the recombination of the dissociated fragments is not a particularly preferred pathway, and is just as possible as any other reaction. This is in line with the findings by \citet{Andersson06} that some of the fragments of photodissociated molecules tend to travel a few molecules' worth away from their site of origin. Thus, the recombination possibility for the fragments is similar to the possibility for reactions with other species in that sublayer.

\section{Results}
\label{res}

As a benchmark reference sheet for the evaluation of calculation results the observational results by \citet{Whittet07} are used. They provide the relative proportions of the major ice components -- H$_2$O, CO, and CO$_2$ -- in samples of interstellar molecular gas towards background stars in Taurus complex of dark clouds. Because the observational studies provide a visual extinction value for each field star, the results are presented in an $A_V$-dependent manner, too. The objects are listed in Table~\ref{tab-obs}. Each of them has an unique $A_V$, and I use their $A_V$ values as an identifier for subsequent results tables. For additional result evaluation, the ice component abundances at threshold $A_V$ -- specified in Table~\ref{tab-obs} -- are used.

The use of data from \citet{Whittet07} allows an evaluation of calculated ice composition that is more detailed than that of other recent dark core models. For example, \citet{Garrod11,Vasyunin13a}; and \citet{Chang14} use a comparison at arbitrary points in time of the simulation run with Elias~16 or the averaged observational results by \citet{Boogert11} and \citet{Oberg11a}. Contrary, the observations of ices towards eight background stars have been used for comparison in the present study, and the $A_V$ of each star was used to tie the observational data to a particular point in time in the model.

The methods and results for the different selective desorption mechanisms were investigated on a case-by-case basis. The Standard model, described in section~\ref{mod}, was used as a matrix, where a single parameter was changed and its effects evaluated. For an easier perception of the meaning of model results, results tables present the calculated-to-observed abundance ratio. The CO:H$_2$O and CO$_2$:H$_2$O ratios are considered (see note~2 below Table~\ref{tab-obs}).
% Table 1
\begin{table*}
\begin{center}
\footnotesize
\caption{The observational data -- $A_V$ and abundance ratios -- used for comparison with calculation results, and the corresponding results of the Standard model. Data from \citet{Whittet07}, Table 1, unless otherwise noted. See \citet{Whittet07} for more details. The calculated abundance of H$_2$O and CO$_2$ in monolayers for their respective $A_{\rm th}$ is given in the lower part of the table.}
\label{tab-obs}
  \begin{tabular}{lllccccc}
\tableline
\tableline
 &  &  &  & \multicolumn{2}{c}{Observations}  & \multicolumn{2}{c}{Standard model} \\
Source ID & Association & $A_V$ & Time\tablenotemark{1}, kyr & $\rm \frac{CO}{H_2O}$,\% & $\rm \frac{CO_2}{H_2O}$,\% & $\rm CO \frac{calc}{obs}$\tablenotemark{2} & $\rm CO_2 \frac{calc}{obs}$ \\
\tableline
043728.2+261024 & Tamura 2 & 6.3 $\pm$ 1.5 & 932 & . . . & 25.0 & . . . & 3.8 \\
042324.6+250009 & Elias 3 & 10 $\pm$ 0.5 & 958 & 20.2 & 19.1 & 1.7 & 3.1 \\
043325.9+261534 & Elias 13 & 11.7 $\pm$ 0.5 & 964 & 12.0 & 16.7 & 3.2 & 3.1 \\
043926.9+255259 & Elias 15 & 15.3 $\pm$ 0.5 & 973 & 27.3 & 16.7 & 1.7 & 2.6 \\
042630.7+243637 &  & 17.8 $\pm$ 1.5 & 976 & 45.1 & 18.3 & 1.1 & 2.2 \\
043213.2+242910 &  & 20.9 $\pm$ 1.5 & 980 & 33.3 & 16.0 & 1.6 & 2.4 \\
044057.5+255413 & Tamura 8 & 21.5 $\pm$ 0.5 & 981 & 24.3 & . . . & 2.2 & . . . \\
043938.9+261125 & Elias 16 & 24.1 $\pm$ 0.5 & 983 & 25.3 & 21.0 & 2.2 & 1.7 \\
\tableline
 & &  &  &  & H$_2$O, ML & CO$_2$, ML & CO, ML \\
H$_2$O threshold\tablenotemark{2} &  & 3.2 $\pm$ 0.1 & 838 &  & 6.7 &  &  \\
CO$_2$ threshold &  & 4.3 $\pm$ 1.0 & 894 &  &  & 12.0 &  \\
CO threshold\tablenotemark{3} &  & 6.8 $\pm$ 1.6 & 938 &  &  &  & 3.5 \\
\tableline
\end{tabular}
	\tablenotetext{1}{Time required to achieve the particular $A_V$ in the model}
 	\tablenotetext{2}{Equal to $\rm \frac{[CO/H_2O]{calc}}{[CO/H_2O]{obs}}$}.
	\tablenotetext{3}{\citet{Whittet01}}
%	\tablenotetext{4}{\citet{Bergin05}}
\end{center}
\end{table*}

\subsection{The Standard model}
\label{r-stand}
%
% Figure 2
\begin{figure}
% \vspace{4cm}
  \hspace{-1cm}
  \includegraphics{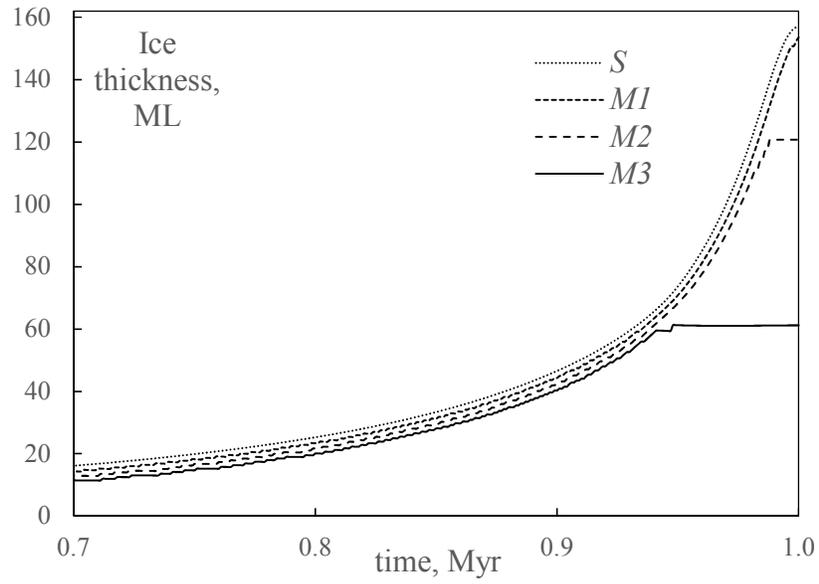}
 \vspace{-15cm}
 \caption{Ice accumulation (total thickness in monolayers) on grains during late cloud evolution for the Standard model. $S$ is the surface layer, $M1$, $M2$, and $M3$ are sublayers 1, 2, and 3, respectively.}
 \label{att-sub}
\end{figure}
%
% Figure 3
\begin{figure}
 \vspace{-13cm}
  \hspace{-2cm}
  \includegraphics{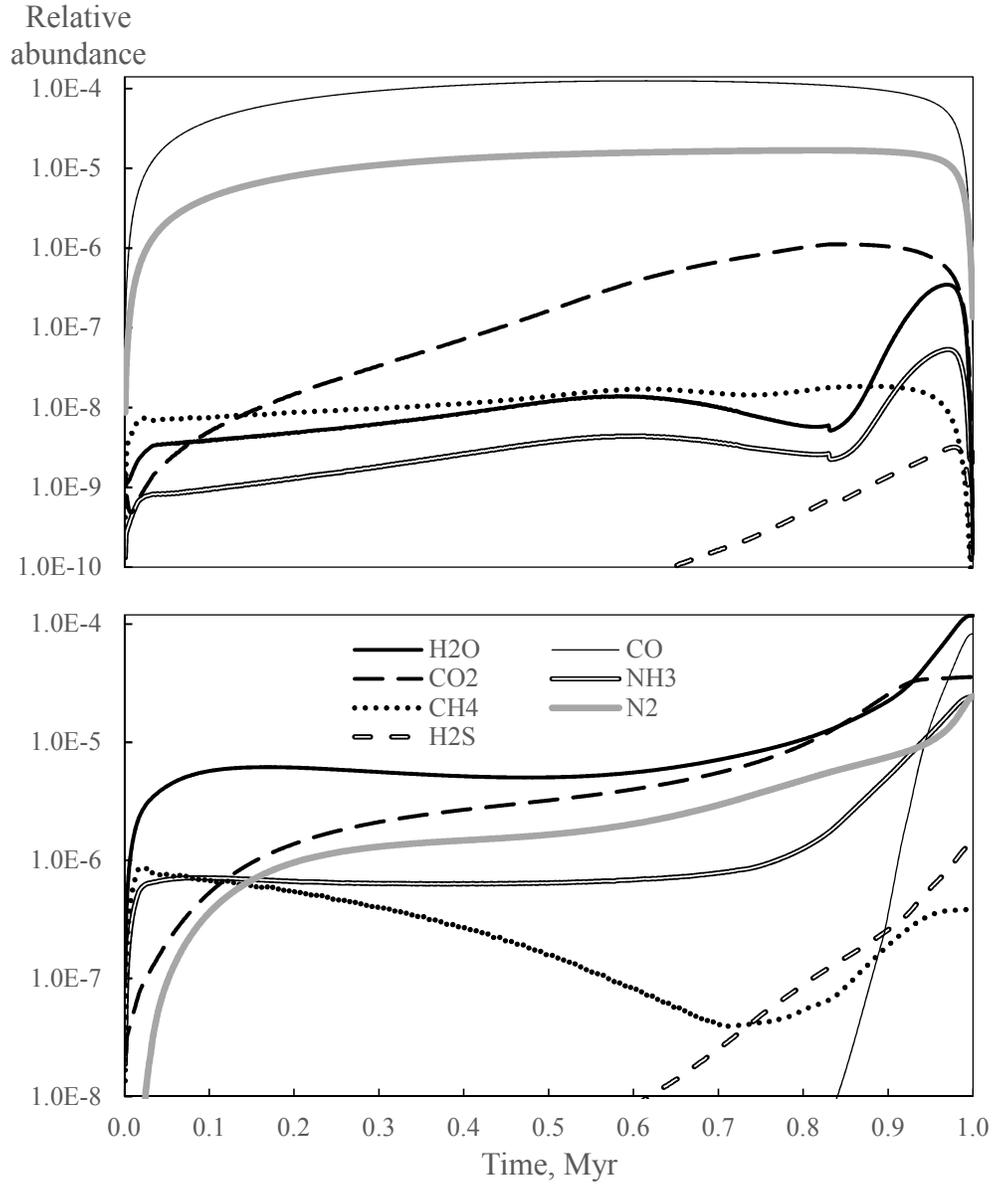}
 \vspace{-2cm}
 \caption{The Standard model: calculated gas-phase (top panel) and ice (bottom panel) abundances of major ice species, relative to hydrogen.}
 \label{att-st}
\end{figure}
%
% Figure 4
\begin{figure}
 \vspace{-13cm}
  \hspace{-2cm}
  \includegraphics{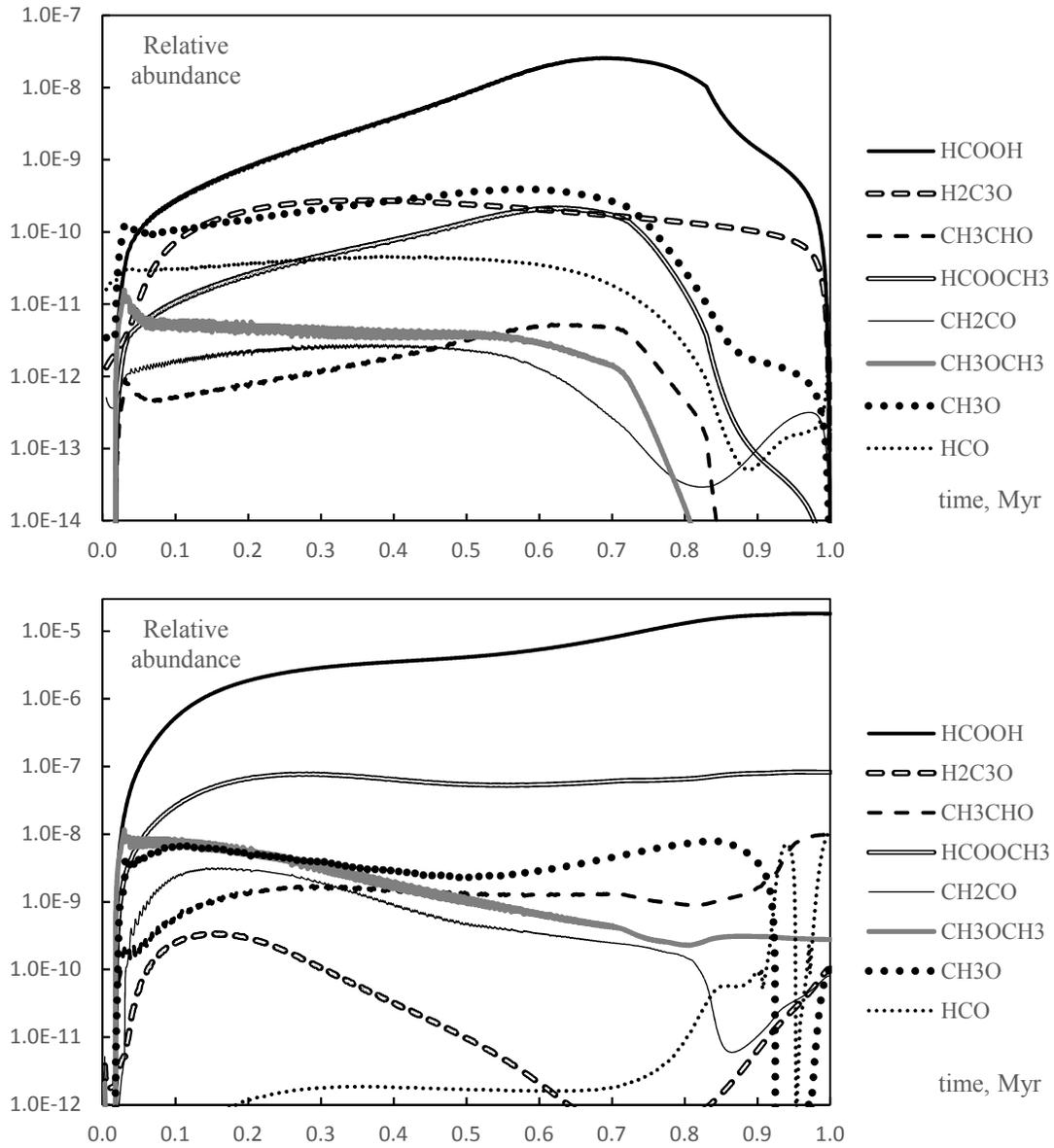}
 \vspace{-2cm}
 \caption{Gas-phase (top panel) and ice (bottom panel) abundances, relative to hydrogen, of the Standard model for organic species observed in quiescent dark clouds. The `tremble' of some abundance curves at early times is an artifact, related to the step-like transition of surface species into sublayer 1.}
 \label{att-st-com}
\end{figure}
Figure~\ref{att-sub} shows the calculated the evolution of ice structure. A slow ice accumulation starts from the earliest stages of the model because the low photodesorption yield is insufficient to keep the grains clean at $A_V\approx2$.

Table~\ref{tab-obs} shows that both, CO and CO$_2$, are significantly overproduced. Such a result was obtained, partially because the calculation results are compared to observational results tied to a particular $A_V$, and not to single averaged values, which is the case in many previous papers. This implies the assumption that the observational sample consists of low-mass prestellar cores. They are in different stages of evolution, or, alternatively, the field stars sample molecular gas in different distances from the center of the core.

The thickness at which photodesorption achieves its maximum efficiency ($B_{\rm ph}$) is approximately 3MLs in this model. Test calculations with different $Y_{\rm ph}$ reveal that a grain can be covered in ice with thickness lower or similar to $B_{\rm ph}$ for very long timescales at $A_V$ lower than the water threshold value. 

Water ice abundance of seven monolayers at $A_{\rm th}$ is significantly more than the single ice layer adopted as a limit for water ice detection by \citet{Garrod11}. The total ice thickness at 3.2mag is 27ML, with HCOOH, CO$_2$, N$_2$, and NH$_3$ being other major species. This result of the Standard model can be consistent with observations. \citet{Whittet01} conclude that water $A_{\rm th}$ is coincident with an increase of grain size, most probably by ice mantle accumulation. Approximately 20 to 30 monolayers of ice are required for a moderate increase of the grain radius by ten per cent. In addition, \citet{Whittet01} allows the possibility that the ice is concentrated in a clump along the line of sight. This would imply that the actual $A_{\rm th}$ is halved, although such an extreme case is unlikely. Both of these considerations permit the existence of water ice at lower $A_V$ or thicker ice at $A_{\rm th}$. Water ice abundance, relative to hydrogen, reaches $10^{-4}$, which is in agreement with observations of molecular clouds \citep{Boogert04,Whittet07}.

Figure~\ref{att-st} shows the evolution of the abundances of major ice species. A comparison with observational data \citep[][Elias 16 with $A_V\approx24.1$]{Gibb04,Oberg11b} reveals that ammonia is overproduced (NH$_3$:H$_2$O=21\%), while methanol is underproduced (CH$_3$OH:H$_2$O=0.1\%). NH$_3$ is synthesized by a surface reaction pathway, similar to that of water, and both compounds have very similar $E_D$. Thus, ammonia is almost inseparable of watery ices in the model, and always has a relatively high abundance. Methanol has a low abundance largely because its synthesis is hampered by barriers in formaldehyde hydrogenation reactions. These effects are unlikely to significantly affect the proportions of H$_2$O, CO, and CO$_2$, although they are important for the chemistry of nitrogen and complex organic molecules (COM).

Early ice accumulation at relatively low $A_V$ results in a phenomenon that can be termed as `photon-dominated ice' (PDI). This means that the (relatively thin) ice layer, formed before a high extinction of the ISRF is reached, experiences intensive irradiation by interstellar photons. PDIs are promoted by low yields for $A_V$-dependent desorption mechanisms -- interstellar photons and reactive desorption. The Standard model (along with the low-yield reactive desorption Model B) is among the most pronounced representations of PDIs described in this paper.

In the model presented here, photoprocessing does not affect significantly the abundances of major species because radicals in ice regenerate these species. However, the production of organic species is greatly enhanced in bulk PDIs. Ice photoprocessing, coupled to non-thermal desorption mechanisms can be a major source of gas-phase COMs in quiescent clouds. During the long-lived diffuse phase ($n_{\rm H}\approx3000$cm$^{-3}$) it produces abundances on the order of $10^{-12}$...$10-^{-9}$ relative to hydrogen for species that recently have been observed in quiescent gas \citep[][see also Table~\ref{tab-com}]{Oberg10,Bacmann12,Cernicharo12}. Figure~\ref{att-st-com} shows the calculated abundances for several of these species.

The gaseous COM-rich evolutionary phase has a length of almost 1Myr in the Standard model. It starts with the formation of the first sublayer at 18kyr and ends at densities higher than 10$^4$cm$^{-3}$. After this period, the rate of species' transition into sublayers is faster than the synthesis of COMs, which takes several steps to be completed. COMs become buried in ice, and are not available for desorption in the near-surface layers. This mechanism inhibits desorption for any complex species during the freeze-out stage, regardless if they are produced in the surface or the mantle.

The Standard model rather poorly represents ice accumulation, and the COM-rich period is significantly shorter for models with effective desorption mechanisms. Because of this, a detailed discussion on the chemistry of COMs has been reserved for the `complete' model (section~\ref{r-com}), which describes overall ice composition more accurately. Earlier papers that investigate the chemistry and reaction networks of COMs include \citet{Garrod06,Garrod08,Belloche09,Laas11,Taquet12,Vasyunin13b,Garrod13a}. The gas-phase organic-rich period has little effect on most carbon-chain species because they are not produced in the sublayers.

\subsection{Photodesorption}
\label{r-pdes}
% Table 2
\begin{table}
 \centering
 \begin{center}
\caption{Photodesorption yields as derived from experimental data.}
\label{tab-ypd}
\begin{tabular}{lll}
\tableline
\tableline & $Y_{\rm isrf}$ & $Y_{\rm crph}$ \\
\tableline
H$_2$O & 2.7E-03 & 1.8E-03\tablenotemark{a} \\
CO\tablenotemark{b} & 5.7E-03 & 3.0E-03 \\
CO$_2$ & 3.5E-03 & 2.3E-03\tablenotemark{c} \\
N$_2$\tablenotemark{b} & 5.5E-03 & 3.0E-03 \\
O$_2$\tablenotemark{d} & 3.3E-03 & 2.6E-03 \\
Other & 3.0E-03 & 2.0E-03 \\
\tableline
\end{tabular}
	\tablenotetext{a}{ \citet{Oberg09a}}
	\tablenotetext{b}{ \citet{Bertin13}}
	\tablenotetext{c}{ \citet{Oberg09b}}
	\tablenotetext{d}{ \citet{Fayolle13}}
\end{center}
\end{table}
%
% Figure 5
\begin{figure}
 \vspace{-10cm}
  \hspace{-2cm}
  \includegraphics{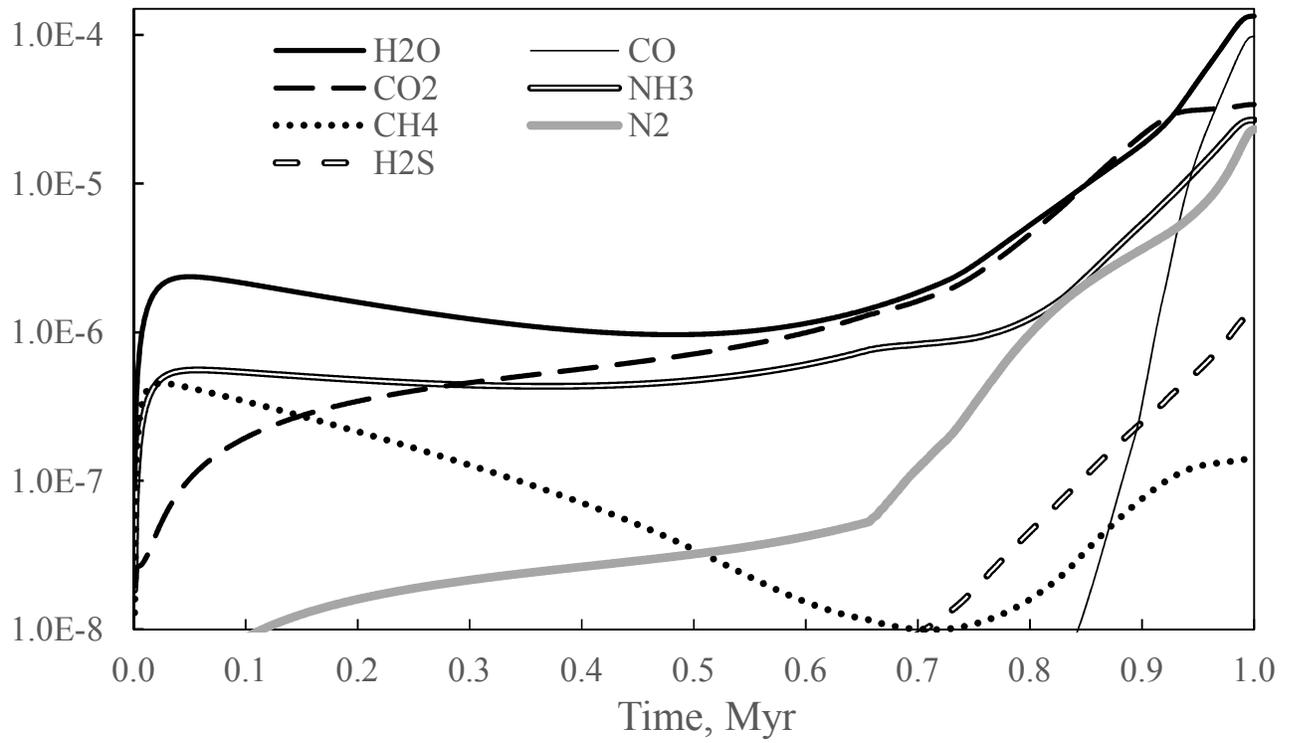}
 \vspace{-10cm}
 \caption{Calculated abundances of major ice species, relative to hydrogen, photodesorption simulation with empiric, $E_D$-dependent yield. Compare with figure~\ref{att-st}.}
 \label{att-pd-emp}
\end{figure}
%
% Table 3
\begin{table}
\begin{center}
\caption{Comparison of observations and calculations with modified photodesorption yields for ice species: abundances of ice species CO and CO$_2$, relative to water, and abundance in monolayers for H$_2$O and CO$_2$ at threshold $A_V$. Photodesorption model A with uniform yield for all species. Models that use experimentally detected photodesorption yields or empiric, $E_D$-dependent yields give very similar results (see text).}
\label{tab-pd}
  \begin{tabular}{lcccccc}
\tableline
\tableline
 & \multicolumn{2}{c}{A} \\
\tableline
$A_V$ & $\rm CO\frac{calc}{obs}$ & $\rm CO_2\frac{calc}{obs}$ \\
6.3 & ... & 3.9 \\
10 & 1.9 & 2.8 \\
11.7 & 3.6 & 2.8 \\
15.3 & 1.8 & 2.3 \\
17.8 & 1.2 & 2 \\
20.9 & 1.7 & 2 \\
21.5 & 2.3 & ... \\
24.1 & 2.3 & 1.5 \\
\tableline
 & \multicolumn{2}{c}{H$_2$O:CO:CO$_2$(ML)} \\
$A_{\rm th}$ & \multicolumn{2}{c}{3.9:3.9:10.2} \\ 
\tableline
\end{tabular}
\end{center}
\end{table}
Desorption by ISRF and secondary photons has been often treated in a non-selective manner in recent papers \citep{Vasyunin13a,Vasyunin13b,Chang14,Gerner14}. \citet{Garrod11} use experimental photodesorption data, obtained for pure species (water, carbon oxides, nitrogen), in their models, which help to explain $A_{\rm th}$ for different ice constituents. These authors adopt similar desorption yields for ISRF and secondary photons.

Experiments have shown that codesorption from icy mixtures may result in similar yield for different species \citep{Oberg09b,Bertin13}, and that photodesorption efficiency is wavelength-dependent \citep{Fayolle11b,Fayolle13,Bertin13}. In order to investigate these factors, three simulations have been run, each with a different approach on photodesorption.

The first simulation made use of experimental data, where possible. For ISRF-induced desorption the data by \citet{Fayolle13} and \citet{Bertin13} was used. These authors, as well as \citet{Oberg09a,Oberg09b}, also provide data for desorption by secondary photons. Molecules with available yields are H$_2$O, CO, CO$_2$, N$_2$, and O$_2$. I could not find recent information on $Y_{\rm isrf}$ for H$_2$O and CO$_2$. From the experimental data, it can be estimated that $Y_{\rm isrf}\approx1.5 Y_{\rm crph}$. For species with no data available, the yields have been estimated. It was assumed that they are slightly higher than water photodesorption yields, and retain the 1.5 ratio. These data are summarized in Table~\ref{tab-ypd}. All experimental results are relevant for a minimum temperature of 15K and, and, for consistency, all the yields used in this model are those for 15K. The yields also assume mantles of infinite thickness. Some thickness-dependent effects have been already included into the model (section~\ref{gasgr}). Only the yields provided by \citet{Bertin13} take into account the codesorption effect (for a CO:N$_2$ mixture).

In the second simulation uniform photodesorption yields for all species were assumed (Table~\ref{tab-ypd}, last row), i.e., codesorption is prevalent.

For the third simulation, a weak, empiric yield dependence on $E_D$ was used:
   \begin{equation}
   \label{res1}
Y_{\mathrm{isrf}} = Q_0 \mathrm{exp}(-E_D/(Q_1+100T)),
   \end{equation}
where $Q_0=0.007$ and $Q_1=4200$. Eq.~(\ref{res1}) approximately matches the experimental data and, by extrapolation, fits the temperature dependence for water photodesorption \citep{Oberg09b} in the relevant temperature interval (6-16K). It was applied for all surface species. As in the first two simulations, $Y_{\rm isrf}=1.5Y_{\rm crph}$.

All of these photodesorption models produce very similar abundance patterns, and an example is shown in figure~\ref{att-pd-emp}. The abundance of carbon oxides in ice, relative to water at different $A_V$ values (Table~\ref{tab-obs}), is very similar for all three simulations, as well as the abundance of these species at $A_{\rm th}$. These numbers do not deviate by more than 5\% for the three simulations. Because of this, the results of the uniform-yield simulation (Model A, see below) have been presented in Table~\ref{tab-pd}, only. The effective desorption yields in real interstellar conditions should be much more selective than experiments indicate in order to introduce significant changes for the H$_2$O:CO$_2$ ratio. Based on current evidence, it can be concluded that, although photodesorption largely regulates molecule accretion on grains under diffuse-cloud conditions, it can be regarded as non-selective because it basically does not affect the relative proportions of ice ingredients. Similarly to the Standard model, models with modified photodesorption yields show a period of a relatively high COM gaseous abundances. Because of a higher $Y_{\rm ph}$, this phase begins much later, at around 0.6Myr.

Based on these results, I conclude that it is unnecessary to use a complex, selective approach for molecule photodesorption from interstellar grains. Thus, the simplest approach with uniform desorption yields (0.002 for secondary and 0.003 for ISRF photons) was chosen as the most appropriate photodesorption model. This is further referenced to as Model A.

\subsection{Reactive desorption}
\label{r-rdes}
%
% Table 4
\begin{table*}
\begin{center}
\caption{Comparison of calculation results. Reactive desorption models B ($\alpha$=0.01), C ($\alpha$=0.03), and D ($\alpha$=0.06).}
\label{tab-rd}
  \begin{tabular}{lcccccc}
\tableline
\tableline
 & \multicolumn{2}{c}{B} & \multicolumn{2}{c}{C} & \multicolumn{2}{c}{D} \\
\tableline
$A_V$ & $\rm CO \frac{calc}{obs}$ & $\rm CO_2 \frac{calc}{obs}$ & $\rm CO \frac{calc}{obs}$ & $\rm CO_2 \frac{calc}{obs}$ & $\rm CO \frac{calc}{obs}$ & $\rm CO_2 \frac{calc}{obs}$ \\
6.3 & . . . & 3.8 & . . . & 4.0 & . . . & 4.2 \\
10 & 1.7 & 3.1 & 1.8 & 3.1 & 2.0 & 3.2 \\
11.7 & 3.2 & 3.1 & 3.5 & 3.2 & 3.7 & 3.2 \\
15.3 & 1.7 & 2.6 & 1.8 & 2.6 & 1.9 & 2.6 \\
17.8 & 1.1 & 2.2 & 1.1 & 2.2 & 1.2 & 2.2 \\
20.9 & 1.6 & 2.4 & 1.7 & 2.3 & 1.8 & 2.4 \\
21.5 & 2.2 & . . . & 2.3 & . . . & 2.5 & . . . \\
24.1 & 2.2 & 1.7 & 2.3 & 1.7 & 2.5 & 1.7 \\
\tableline
 & \multicolumn{2}{c}{H$_2$O:CO:CO$_2$ (ML)} & \multicolumn{2}{c}{H$_2$O:CO:CO$_2$ (ML)} & \multicolumn{2}{c}{H$_2$O:CO:CO$_2$ (ML)} \\
$A_{\rm th}$ & \multicolumn{2}{c}{6.9 : 3.5 : 12.1} & \multicolumn{2}{c}{5.6 : 3.6 : 11.5} & \multicolumn{2}{c}{4.5 : 3.7 : 10.9} \\
\tableline
\end{tabular}
\end{center}
\end{table*}
Reactive desorption was included in the model following the approach by \citet{Garrod07}, based on the Rice-Ramsperger-Kessel (RRK) theory. For clarity, a full description is included here. The fraction of reactions resulting in product desorption is
   \begin{equation}
   \label{res2}
f_\mathrm{rd}= \frac{\alpha P_\mathrm{RRK}}{1 + \alpha P_\mathrm{RRK}},
   \end{equation}
where $\alpha=\nu_0 / \nu_s$ is ``the ratio of the surface–molecule bond frequency to the frequency at which energy is lost to the grain surface'' \citep{Garrod07}. The probability $P_\mathrm{RRK}$ for an energy, larger than $E_D$, to be present in the molecule-surface bond depends on the total energy $E_\mathrm{reac}$ released in an exothermic reaction:
   \begin{equation}
   \label{res3}
P_\mathrm{RRK}= \left(1 - \frac{E_D}{E_\mathrm{reac}}\right)^{s-1},
   \end{equation}
where $s=3N-5$ and $N$ the number of atoms in the most complex product molecule ($s=2$ if $N=2$).

$E_\mathrm{reac}$ is equal to negative reaction standard enthalpy, $\Delta H_r^0$. In the model, the latter was calculated for each reaction, using the gas-phase standard enthalpies of formation $\Delta H_f^0$ of reactants and products. Standard enthalpies were chosen, because more species have thermochemical data available for 273K than for 0 or 10K, while the changes are relatively insignificant. Similarly, enthalpy, not Gibbs energy, was used because entropy is of little importance at the low temperatures considered in the model.

In order to facilitate the implementation of the reaction-specific reactive desorption into the program, only two cases were considered -- either that all products desorb into the gas or that all products remain on the surface. This is the original approach used in the ALCHEMIC code. Consequently, the sum of product desorption energies were used instead of $E_D$ in Eq.~(\ref{res3}). $N$ used in the same equation was assumed to be the number of atoms in the most complex species among the products. Although not entirely physically justified, these approximations generally retain the unique selectivity of the reactive desorption process. Because of the dependence on two variables -- $E_\mathrm{reac}$ and $E_D$ -- reactive desorption is different from other mechanisms, which have $E_D$ as their only variable, unique for each species.

The use of actual $E_\mathrm{reac}$ and $E_D$ values for each reaction in Eq.~(\ref{res3}) shows that the assumption by \citet{Garrod07} that $P_\mathrm{RRK}\approx1$, and thus $f_\mathrm{rd}\approx\alpha$, is seldom applicable for surface reactions. If $\alpha$ is taken between 0.01 and 0.1, less than a quarter of all reactions have $f_\mathrm{rd}>0.5\alpha$. Usually, these are oxidation-reduction reactions with high exothermicity between light species. Atom exchange reactions involving heavy, complex molecules tend to have insignificant desorption probabilities. The changes implemented in this model mean that reactive desorption is a truly selective desorption mechanism, with $\alpha$ as its only poorly-known parameter \citep[see][for a more detailed discussion]{Garrod07}. Calculations were performed with $\alpha$ values in the range 0-0.2, and Table~\ref{tab-rd} shows results for $\alpha$=0.01 (Model B), 0.03 (C), and 0.06 (D).

It can be seen that in terms of carbon oxide abundances relative to water, Model~B is the best choice, although it still has a significant overproduction of CO$_2$. The low efficiency of reactive desorption in Model~B means that the total rate of desorption is actually lower than in the Standard model. The higher $\alpha$ values for Models C and D result in an even higher CO$_2$:H$_2$O ratio. All the formation reactions of these two molecules are highly exothermic and result in high desorption efficiency ($f_{\rm rd}>0.7 \alpha$). Because water forms in two steps, while CO$_2$ forms in two parallel one-step reactions, water is desorbed more effectively than carbon dioxide. These results mean that the model presented here does not place new constraints on $\alpha$, although lower values may favor a CO$_2$:H$_2$O ratio that is slightly closer to observations, as shown in Table~\ref{tab-rd}. CO is formed by gas-phase reactions and is practically unaffected by reactive desorption.

Selective reactive desorption favors the formation of COMs in ice, with COM abundances of Model~B being higher than those in the Standard model results. This is because the high $E_D$ of COMs combined with often low $E_{\rm reac}$ result in typically very low desorption efficiency for most reactions involving organic species. However, at high $\alpha$ values the ice does not accumulate until late times, and the length of the COM-rich cloud evolutionary phase is greatly reduced, if $\alpha\approx0.1$. Such high $\alpha$ values are not supported by current theoretical evidence \citep{Garrod07}.

\subsection{Desorption by H~+~H reaction on grains}
\label{r-hfdes}

The rate coefficient for desorption of ice molecules following an exothermic H$_2$ formation reaction on grain surface \citep[indirect reactive desorption, H$_2$FD for short,][]{Willacy94} is proportional to H$_2$ formation rate per grain per adsorption site:
    \begin{equation}
   \label{res4}
k_\mathrm{H_2fd}= R_\mathrm{acc, H} \times f_\mathrm{H_2fd} \times \left[\frac{[\mathrm H]}{n_g N_s}\right]^2, \mathrm s^{-1},
   \end{equation}
where $R_\mathrm{acc, H}$, s$^{-1}$ is the rate of accretion for H atoms onto grains and $f_\mathrm{H_2fd}$ is an efficiency parameter -- the number of desorbed molecules per formed H$_2$ molecule. The last term in Eq.~(\ref{res4}) is the quadratic abundance of gaseous H atoms per grain per adsorption site. Similarly to photodesorption and evaporation, H$_2$FD can affect sublayer species, if the number of overlying MLs does not exceed two. According to Eq.~(\ref{res4}), indirect reactive desorption considers \emph{any} surface exothermic reaction involving H, with the formation of H$_2$ being the dominant energy source for this mechanism.

No experimental or quantum-chemical data on the possible value of $f_{\rm H_2fd}$ were found in the literature. The only estimates of $f_{\rm H_2fd}$ were found to be provided by astrochemical modeling, e.g., \citet{Roberts07}. These values are not used here. The physical model and methods for H$_2$FD rate calculation differ significantly in \citet{Roberts07} and the present study.

For estimating for $f_{\rm H_2fd}$, a best-fit method was used. This means that simulations with a range of $f_{\rm H_2fd}$ values were performed until the closest possible agreement with the observations by \citet{Whittet07} was achieved.

For the selectivity of H$_2$FD, the approach developed by \citet{Duley93} and \citet{Roberts07} has been used. This means that all species are desorbed with an equal efficiency $f_{\rm H_2fd}$, up to a certain adsorption energy threshold, $E_{\rm th}$. This is probably the simplest possible form for a selective desorption mechanism. H$_2$FD is poorly constrained, and I explore cases with three $E_{\rm th}$ values, as well as empiric equations for $f_{\rm H_2fd}$. The latter have been developed taking into account the results obtained by calculations with the $E_{\rm th}$ approach.

The three threshold cases are with $E_D\leq$1210K, $E_D<$2600K, and $E_D<$6000K. The first value is the one used by \citet{Roberts07}. For water and carbon oxides this means that only CO desorption is enabled in the first case, CO and CO$_2$ may desorb in the second case, and all three components desorb in the third case. The obtained calculated-to-observed ice relative abundance ratios CO and CO$_2$ are shown in Table~\ref{tab-hdthr}. Additionally, two models with a more complex approach on $E_D$-dependent desorption efficiency have been investigated.

\subsubsection{1210K desorption energy threshold}
\label{r-1210}
%
% Table 5
\begin{table*}
\begin{center}
\caption{Comparison of calculation results. Indirect desorption by H+H $\longrightarrow$ H$_2$ reaction for molecules with adsorption energy threshold approach: $E_D\leq1210$K and $f_\mathrm{H_2fd}=4\times10^{-6}$ (Model E), $E_D<2600$K and $f_\mathrm{H_2fd}=2\times10^{-6}$ (Model F), and $E_D<6000$K and $f_\mathrm{H_2fd}=2\times10^{-6}$ (Model G).}
\label{tab-hdthr}
  \begin{tabular}{lcccccc}
\tableline
\tableline
 &  \multicolumn{2}{c}{E} &  \multicolumn{2}{c}{F} &  \multicolumn{2}{c}{G} \\
\tableline
$A_V$ &  $\rm CO \frac{calc}{obs}$ &  $\rm CO_2 \frac{calc}{obs}$ &  $\rm CO \frac{calc}{obs}$ &  $\rm CO_2 \frac{calc}{obs}$ &  $\rm CO \frac{calc}{obs}$ &  $\rm CO_2 \frac{calc}{obs}$ \\
6.3 &  . . . & 4.6 &  . . . & 0.9 &  . . . & 1.5 \\
10 & 0.5 & 4.2 & 0.9 & 0.9 & 2.9 & 1 \\
11.7 & 1.1 & 4.3 & 1.8 & 0.9 & 5.1 & 1.1 \\
15.3 & 0.8 & 3.5 & 1.1 & 0.9 & 2.3 & 1 \\
17.8 & 0.5 & 2.9 & 0.7 & 0.8 & 1.4 & 0.9 \\
20.9 & 0.9 & 3.1 & 1.2 & 0.9 & 2 & 1 \\
21.5 & 1.3 &  . . . & 1.7 &  . . . & 2.8 &  . . . \\
24.1 & 1.4 & 2.2 & 1.8 & 0.7 & 2.7 & 0.7 \\
\tableline
 &  \multicolumn{2}{c}{H$_2$O:CO:CO$_2$ (ML)} &  \multicolumn{2}{c}{H$_2$O:CO:CO$_2$ (ML)} &  \multicolumn{2}{c}{H$_2$O:CO:CO$_2$ (ML)} \\
$A_{\rm th}$ &  \multicolumn{2}{c}{6.4 : 1.1 : 11.3} &  \multicolumn{2}{c}{7.1 : 1.9 : 2.8} &  \multicolumn{2}{c}{0.5 : 1.5 : 0.7} \\
\tableline
\end{tabular}
\end{center}
\end{table*}
For a threshold energy of 1210K, calculations with $f_\mathrm{H_2fd}$ values in the range $10^{-6}-10^{-4}$ were performed. The most likely value was found to be near $4\times10^{-6}$. Results of Model E with $f_\mathrm{H_2fd}=4\times10^{-6}$ are shown in Table~\ref{tab-hdthr}. The model produces a rather good agreement with observations for the CO:H$_2$O ratio. Although carbon dioxide is synthesized via CO, the CO$_2$:H$_2$O ratio remains almost unchanged from in the Standard model. $f_\mathrm{H_2fd}$ is roughly inversely proportional to the abundance of gas-phase atomic H.

\subsubsection{2600K desorption energy threshold}
\label{r-2600}

The application of higher values of $E_{\rm th}$ is an extrapolation of the considerations discussed by \citet{Roberts07} and \citet{Duley93}. This is partially justified by the fact that lower values of $f_{\rm H_2fd}$ are required by the approach used here. A threshold $E_D<2600$K means that H$_2$FD affects both, CO and CO$_2$, and that they have equal desorption yields. With such an approach it is possible to achieve a much closer match for observation-modeling results of H$_2$O:CO$_2$:CO$_2$ ice abundance ratio than for all previous methods. Namely, the calculated-to-observed relative abundance ratio varies only within a factor of two for CO and CO$_2$. The CO:H$_2$O ratio has a higher spread. Slightly lower efficiencies than those in the $E_{\rm th}=1210$K model are required to achieve the best agreement with observations. Table~\ref{tab-hdthr} shows the results of Model F with a 2600K threshold and the best-fit value for $f_\mathrm{H_2fd}$ of $2\times10^{-6}$.

These results show that a H$_2$FD mechanism that desorbs CO and CO$_2$, and excludes water, can bring a solution for the $A_V$-dependence of the relative abundances for major ice constituents. However, a threshold of 2600K either underproduces CO$_2$ or overproduces CO ice. This behavior has the element of $E_D$-dependence. In order to further investigate this, an empiric H$_2$FD approach is presented in the next subsection.

I would like to note that, similarly to all other models, CO ice abundance at $A_{\rm th}$ (in ML) is significantly lower to respective H$_2$O and CO$_2$ values. Nevertheless, the $A_V$-dependent results middle column in Table~\ref{tab-hdthr} clearly show that CO is the most overproduced at $A_{\rm th}$ of the three major ice species. This suggests that the chosen $A_{\rm th}$ value might be inaccurate. We shall return to this issue in the Conclusions section.

\subsubsection{Other approaches on H$_2$FD efficiency}
\label{hf-other}
%
% Table 7
\begin{table*}
\begin{center}
\caption{Comparison of calculation results. Indirect reactive desorption efficiency calculated with the use of the RRK theory (Model H), and by assuming a thermal desorption approach (Model I).}
\label{tab-hdciti}
  \begin{tabular}{lcccccc}
\tableline
\tableline
 & \multicolumn{2}{c}{H} & \multicolumn{2}{c}{I} \\
\tableline
$A_V$ & $\rm CO \frac{calc}{obs}$ & $\rm CO_2 \frac{calc}{obs}$ & $\rm CO \frac{calc}{obs}$ & $\rm CO_2 \frac{calc}{obs}$ \\
6.3 & . . .  & 1.2 & . . . & 1.2 \\
10 & 0.5 & 1.1 & 0.8 & 1.0 \\
11.7 & 1.0 & 1.2 & 1.5 & 1.1 \\
15.3 & 0.7 & 1.1 & 0.9 & 1.0 \\
17.8 & 0.5 & 1.0 & 0.6 & 0.9 \\
20.9 & 0.8 & 1.1 & 1.0 & 1.0 \\
21.5 & 1.2 & . . . & 1.5 & . . . \\
24.1 & 1.3 & 0.8 & 1.6 & 0.7 \\
\tableline
 & \multicolumn{2}{c}{H$_2$O:CO:CO$_2$ (ML)} & \multicolumn{2}{c}{H$_2$O:CO:CO$_2$ (ML)} \\
$A_{\rm th}$ & \multicolumn{2}{c}{7.0 : 1.0 : 4.6} & \multicolumn{2}{c}{3.9 : 1.3 : 2.6} \\
\tableline
\end{tabular}
\end{center}
\end{table*}
It is possible that H$_2$FD affects molecules with an even higher $E_D$. Calculations were performed for a threshold energy of 6000K, which means that H$_2$FD was attributed to water and ammonia. The calculations produced an excess of CO, CO$_2$ in an very thin ice (Model G in Table~\ref{tab-hdthr}).

Selectivity achieved by assuming a desorption energy threshold is a rather crude approximation. Two cases were considered, where the desorption efficiency is estimated in a more targeted manner. First, the RRK theory, which has been used in the case of direct reactive desorption (section~\ref{r-rdes}), was applied also for indirect desorption (Model H). This was based on an assumption that indirect desorption occurs, when the newly formed and highly excited H$_2^*$ molecule transits energy to (or `kicks') a nearby molecule. Taking into account the calculation results with energy thresholds, the calculation parameters are chosen so that they produce $f_\mathrm{H_2fd}$ close to $2\times10^{-6}$ for CO and CO$_2$ and are negligible for H$_2$O. This means that $f_\mathrm{H_2fd}$ is calculated by Eqs.~(\ref{res2}) and (\ref{res3}) with $\alpha=6\times10^{-6}$ and $E_\mathrm{reac}=6\times10^3$K. Only molecules with $E_D<E_\mathrm{reac}$ can be considered. These parameters have been applied for H$_2$FD, which was included the Standard model, described in Sect~\ref{mod}.

Another approach is to treat H$_2$FD as evaporation, i.e., to assume that the energy transit from H$_2^*$ results in an immediate temperature increase for the nearby molecule (Model I). The desorption efficiency is calculated with an Arrhenius-type equation $f_\mathrm{H_2fd}=Q_2 \mathrm{exp}(-E_D/(Q_3)$. The parameter $Q_2=7\times10^{-6}$ can be expressed as
   \begin{equation}
   \label{res5}
Q_2= \nu_0 t(Q_3) P_\mathrm{H_2fd},
   \end{equation}
where $t(Q_3)$, s, is the time spent for the nearby molecule at temperature $Q_3=1.7\times10^3$K, and $P_\mathrm{H_2fd}$ is the probability for the energy transfer from H$_2^*$. Regardless of this, the approach used here can be viewed as an empirical derivation, only. Quantum effects play a significant role on the level of a single molecule.

Table~\ref{tab-hdciti} shows that models H and I produce a reasonably good fit to the observational results by \citet{Whittet07}. All the calculated CO:H$_2$O and CO$_2$:H$_2$O ice abundance ratios fall within a $\pm$50\% margin of the observed values. This is better than Model F with $E_{\rm th}=$2600K, where it was found to be impossible to achieve a similar degree of agreement for CO and CO$_2$ simultaneously (Table~\ref{tab-hdthr}). Additionally, Models H and I produce a much greater similarity for H$_2$O and CO$_2$ abundances at their respective threshold $A_V$.

\subsubsection{Summary for H$_2$FD}
\label{hf-sum}

From the results of Models E through I (Tables \ref{tab-hdthr} and \ref{tab-hdciti}) it can be seen that the best agreement with observations for the H$_2$FD mechanisms is achieved by models with (a) effective desorption of CO and CO$_2$, (b) no or very low desorption efficiency for H$_2$O, and (c) CO desorbed slightly more effectively than CO$_2$. The sequence in CO$>$CO$_2>>$H$_2$O for desorption efficiency implies a dependence on the adsorption energy $E_D$. A semi-empirical (Model H) or empirical (Model I) approach produces an almost perfect fit to the observational H$_2$O:CO:CO$_2$ ratio.

It has to be emphasized that the rate of H$_2$FD is highly dependent on the gas phase abundance of atomic hydrogen, which, in turn, is determined by the the chosen physical parameters of the model (gas density and $A_V$). Other assumed properties, such as the number of surface MLs or desorption from subsurface layers, can be of importance, too. The obtained $f_\mathrm{H_2fd}$ values are valid for a model with $n(\mathrm{H})\approx0.01 n(\mathrm{H}_2)$ during the long quiescent cloud phase with $n_H\approx3000$cm$^{-3}$. $f_\mathrm{H_2fd}$ is roughly inversely proportional to the relative abundance of H. An atomic hydrogen abundance of 0.001 relative to H$_2$ results in that a model with $E_{\rm th}=2600$K with $f_{\rm rd}=4\times10^{-5}$ is the model that produces the best agreement with observations. The parameters $\alpha$ and $Q_2$ for Models H and I, respectively, also scale inversely to $n$(H)$^2$. The major finding is that a CO$\approx$CO$_2>>$H$_2$O sequence in desorption efficiency is required for the reproduction of observational results and that H$_2$FD is a likely candidate mechanism, which could exhibit such a selectivity.

A general trend for indirect reactive desorption models is that there is relatively more CO$_2$ than CO at low $A_V$ and vice versa at high $A_V$ values (see Tables \ref{tab-hdthr} and \ref{tab-hdciti}). This is a tentative trend that shows an inverse dependence on $E_D$. Current experimental results suggest that CO has the higher desorption yield for ISRF photons (Table~\ref{tab-pd}), while direct reactive desorption significantly hampers the accumulation of water. Thus, this apparent discrepancy cannot be removed by adjusting the efficiency of other desorption mechanisms examined in this paper. It is possible that the trend arises from discrepancies in the treatment of H$_2$FD or gas-phase chemistry.

There is a lack of experimental evidence or detailed quantum calculations on exothermic reactions resulting in the desorption of nearby molecules. Currently, it seems that the feasible $f_\mathrm{H_2fd}$ values used in the model do not contradict existing views on reactive desorption. For typical atomic H abundances in dark cores, the efficiency for H$_2$FD is two to four orders of magnitude lower than efficiency of direct reactive desorption. The significance of this mechanism lies in the sheer number of hydrogen atoms sticking to and meeting on grain surfaces. Because of the extremely low efficiency, this process can be hard to verify experimentally. An experiment with a beam of atomic H aimed to an inert cryogenic ice layer, e.g., N$_2$, would be advisable.

\subsection{The complete model}
\label{r-compl}
%
% Table 8
\begin{table*}
\begin{center}
\footnotesize
\caption{Comparison of calculation results the complete model, variants J, K, L, and M (see text and Table~\ref{tab-mod} for an explanation).}
\label{tab-compl}
\tabcolsep=0.11cm
  \begin{tabular}{lcccccccc}
\tableline
\tableline
 & \multicolumn{2}{c}{J} & \multicolumn{2}{c}{K} & \multicolumn{2}{c}{L} & \multicolumn{2}{c}{M} \\
\tableline
$A_V$ & $\rm CO \frac{calc}{obs}$ & $\rm CO_2 \frac{calc}{obs}$ & $\rm CO \frac{calc}{obs}$ & $\rm CO_2 \frac{calc}{obs}$ & $\rm CO \frac{calc}{obs}$ & $\rm CO_2 \frac{calc}{obs}$ & $\rm CO \frac{calc}{obs}$ & $\rm CO_2 \frac{calc}{obs}$ \\
6.3 & . . . & 1.0 & . . . & 1.3 & . . . & 1.2 & . . . & 0.9 \\
10 & 1.0 & 0.9 & 0.5 & 1.1 & 0.8 & 1.0 & 0.4 & 0.8 \\
11.7 & 2.0 & 0.9 & 1.1 & 1.2 & 1.6 & 1.0 & 0.8 & 0.8 \\
15.3 & 1.1 & 0.9 & 0.7 & 1.0 & 0.9 & 0.9 & 0.4 & 0.6 \\
17.8 & 0.8 & 0.8 & 0.5 & 0.9 & 0.6 & 0.8 & 0.3 & 0.5 \\
20.9 & 1.2 & 0.9 & 0.8 & 1.0 & 1.0 & 0.9 & 0.4 & 0.5 \\
21.5 & 1.7 & . . . & 1.2 & . . . & 1.5 & . . . & 0.6 & . . . \\
24.1 & 1.8 & 0.7 & 1.3 & 0.8 & 1.6 & 0.7 & 0.6 & 0.4 \\
\tableline
 & \multicolumn{2}{c}{H$_2$O:CO:CO$_2$ (ML)} & \multicolumn{2}{c}{H$_2$O:CO:CO$_2$ (ML)} & \multicolumn{2}{c}{H$_2$O:CO:CO$_2$ (ML)} & \multicolumn{2}{c}{H$_2$O:CO:CO$_2$ (ML)} \\
$A_{\rm th}$ & \multicolumn{2}{c}{3.5 : 1.9 : 2.4} & \multicolumn{2}{c}{3.6 : 1.0 : 3.8} & \multicolumn{2}{c}{2.5 : 1.3 : 2.2} & \multicolumn{2}{c}{3.7 : 0.4 : 1.6} \\
\tableline
\end{tabular}
\end{center}
\end{table*}
%
%
% Figure 6
\begin{figure}
%\epsscale{.70}
% \vspace{4cm}
%  \hspace{-1cm}
%  \includegraphics{att-sub-r.eps}
  \includegraphics[width=15.0cm]{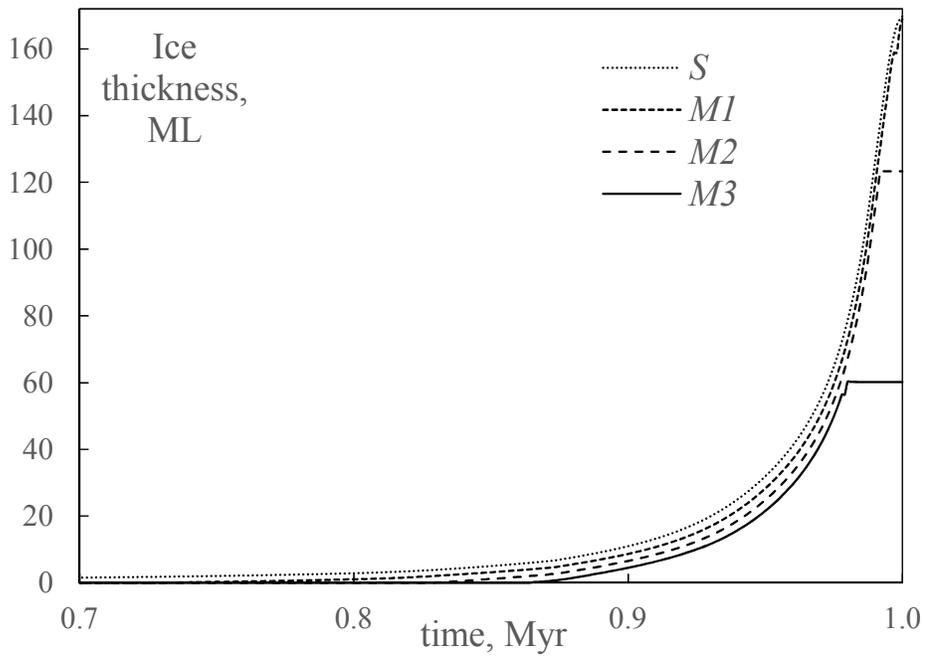}
 \vspace{-5cm}
 \caption{Ice accumulation on grains during late cloud evolution for the complete Model J (cf. figure~\ref{att-sub}).}
 \label{att-sub-r}
\end{figure}
%
% Figure 7
\begin{figure}
 \vspace{-13cm}
  \hspace{-3cm}
  \includegraphics{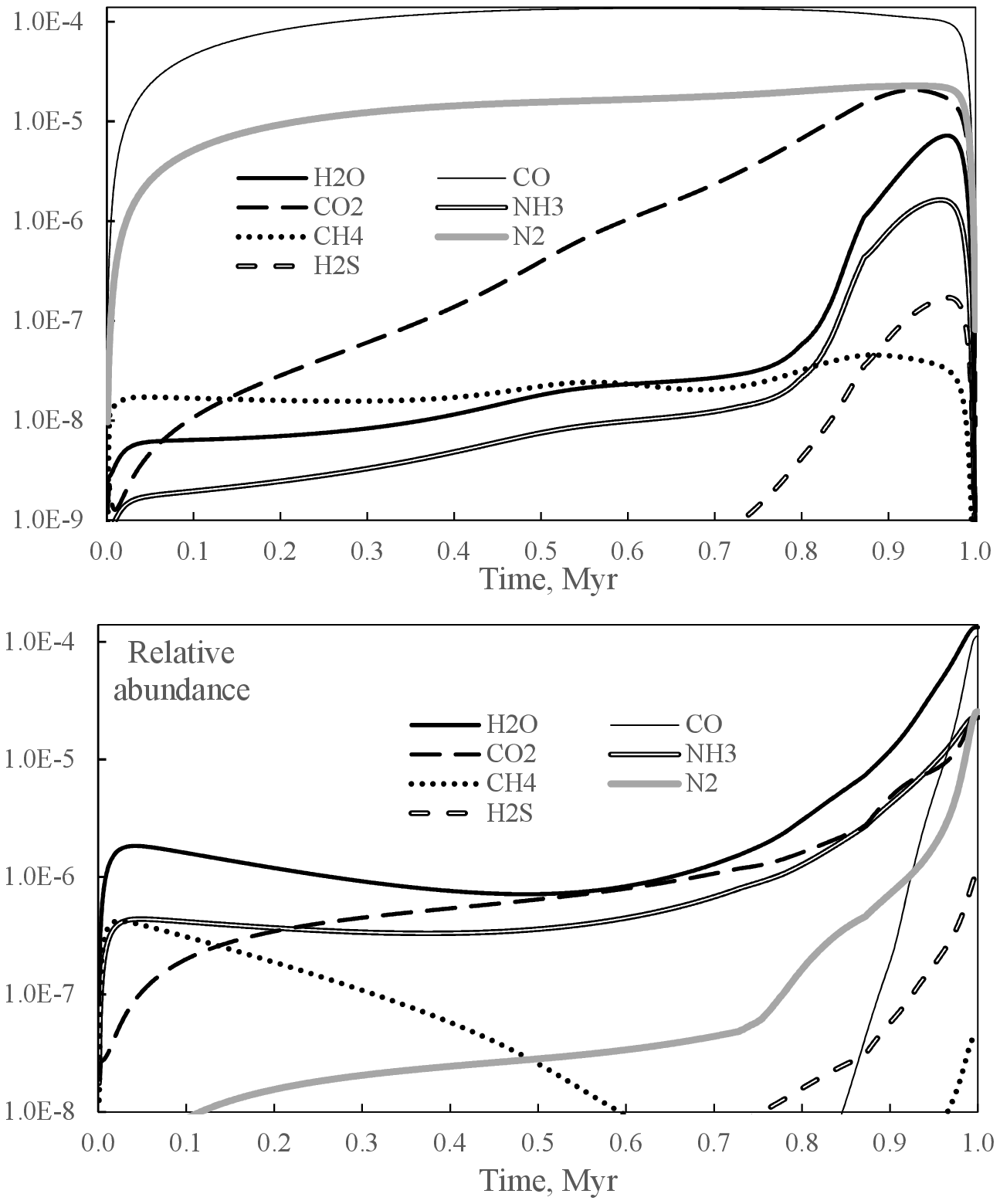}
 \vspace{-2cm}
 \caption{Gas-phase (top panel) and ice (bottom panel) abundances, relative to hydrogen, of major ice species for the complete Model L.}
 \label{att-compl-r}
\end{figure}
Nine derivations of the Standard model have been presented so far, each focusing on the efficiency of a single desorption mechanism with different parameters. I continue with combining the mechanisms in order to create a model that fully and adequately treats molecule desorption during the ice formation epoch in molecular cloud cores.

Evidently, H$_2$FD has the most significant effect on the relative abundances of major ice constituents. The application of H$_2$FD (section~\ref{r-hfdes}) produce calculation results within $\pm$50\% margin of the observed values for CO and CO$_2$ relative abundances in ice. Its efficiency parameters have to be carefully chosen. Desorption by ISRF photons is relatively well constrained and largely regulates the ice accumulation and thickness of ices during the early stages. Reactive desorption has the general effect of favoring small molecules that form in highly exothermic reactions involving free atoms. Its efficiency is governed by $\alpha$, a parameter which is known probably within an order of magnitude.

A general result for all models is that, in terms of this paper, a higher desorption efficiency produce thinner ices at early stages and thicker ices (i.e. more ice molecules) at the end of the integration time. This is because subsurface photoprocessing efficiently produces multi-atom species -- including COMs -- in ice. The abundances of these molecules are lower, if ices accumulate later, and are exposed to ISRF photons for a shorter time-scale. This means that more atoms are concentrated into simple two- or three-atomic species, which means a higher number of molecules. The Standard model produces a final ice thickness of 160ML, while the combined models produce approximately 170ML.

In order to establish a complete, functional model, calculations with various combinations of model cases A through I have been performed. In the complete model, all the three different approaches on photodesorption produce abundances of major species that differ only by hundredths. These are experimentally detected yields, uniform yields for all species, or an empirical relation mimicking the experimental results. The preferred choice here is the simplest case with an uniform desorption yield for all species (Model A).

For reactive desorption, an increase of $\alpha$ in the range of 0.01-0.06 results in higher proportion of CO and lower proportion of CO$_2$ for all $A_V$ values (Table~\ref{tab-rd}). It has the beneficial effect of reducing the abundance of ammonia ice, which is overproduced in all models. All the abundance changes are within a few per cent at most. A more pronounced effect on H$_2$O, CO, and CO$_2$ abundances at $A_{\rm th}$ can be observed (Table~\ref{tab-rd}). 0.03 and 0.01 have been used before as values for $\alpha$ \citep{Garrod07,Garrod11}. Both of these values are appropriate for the current model.

For the indirect reactive desorption mechanism an obvious choice is the empiric approach of Models H and I. A model with $E_{\rm th}=$2600K produces a slightly worse match with observations (section~\ref{hf-sum}). Other H$_2$FD threshold models fail, because a similar desorption efficiency for CO and CO$_2$ and limited or no desorption for H$_2$O is a requirement for reproduction the observed carbon-oxide-to-water proportions. Additionally, Models H and I are probably physically more understandable than the single energy threshold approach for models E to G.

Table~\ref{tab-compl} shows the results of the three most important combined models. The chosen parameters are (1) uniform photodesorption yields ($Y_{\rm isrf}=3\times10^{-3}$ and $Y_{\rm crph}=2\times10^{-3}$); (2) for reactive desorption $\alpha=3\times10^{-2}$; and, (3) three variants for indirect reactive desorption, $E_D$ threshold of 2600K (Model J), H$_2$FD efficiency calculated by the RRK theory (Model K), and H$_2$FD calculated empirically as for a thermal process (Model L).

Out of all these simulations, Model K yields a $\pm$50\% agreement with observations for relative proportions of carbon oxides and water. Model L produces a very similar result. Both these models also produce similar H$_2$O and CO$_2$ abundances that differ by no more than 15\% at their respective $A_{\rm th}$. This can be regarded as an additional positive verification of results. Figs.~\ref{att-sub-r} and \ref{att-compl-r} show the main results of Model L -- the evolution of ice sublayer thickness and the abundances of major ice species, respectively.

All the calculations described above are for a low-mass molecular core, whose $A_V$ increases slowly. In order to provide a simple test for conditions relevant for massive cores, results are presented for Model~M with a contraction time of 0.2Myr, i.e., five times shorter than other simulations. The evolution of other physical parameters from start to end was retained as described in section~\ref{phys}. Model K set of desorption mechanisms was employed in this simulation. As shown in Table~\ref{tab-compl}, Model~M produces up to two times lower abundances of carbon oxides relative to water, when compared to other versions of the complete model. If we were to obtain a better agreement, this discrepancy could be easily corrected by adjusting the efficiency of H$_2$FD.

A likely explanation for these differences is generally lower CO and CO$_2$ ice abundances toward high-mass objects that experience rapid contraction and darkening. Observations of protostars indicate that this indeed may be the case, with high-mass objects having roughly a half of the carbon oxide inventory observed in low- or intermediate-mass objects \citep{Gibb04,Oberg11a}. The abundances of solid H$_2$O:CO and H$_2$O:CO$_2$ ratios for Models L and M approximately agree to these observations. Thus, it can be said that the presented conclusions on selective desorption mechanisms are valid for a wider range of prestellar objects, not just low-mass cores.

\subsubsection{The chemistry of COMs}
\label{r-com}
%
% Figure 8
\begin{figure}
 \vspace{-10cm}
  \hspace{-2cm}
  \includegraphics{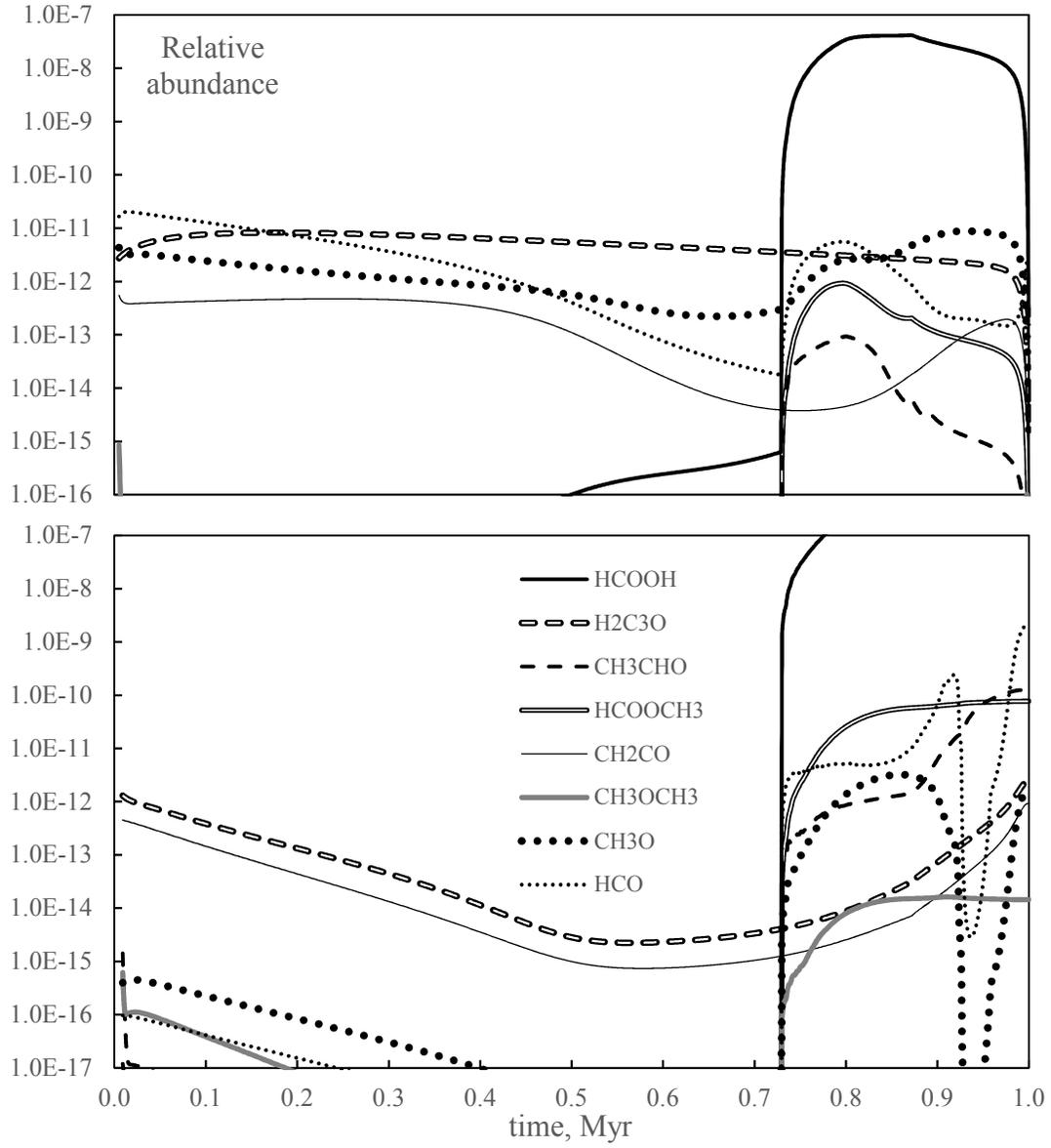}
 \vspace{-5cm}
 \caption{Gas-phase (top panel) and ice (bottom panel) abundances, relative to hydrogen, of the complete Model L for organic species observed in quiescent dark clouds.}
 \label{att-r-com}
\end{figure}
%
% Figure 9
\begin{figure}
 \vspace{-8cm}
  \hspace{-3cm}
  \includegraphics{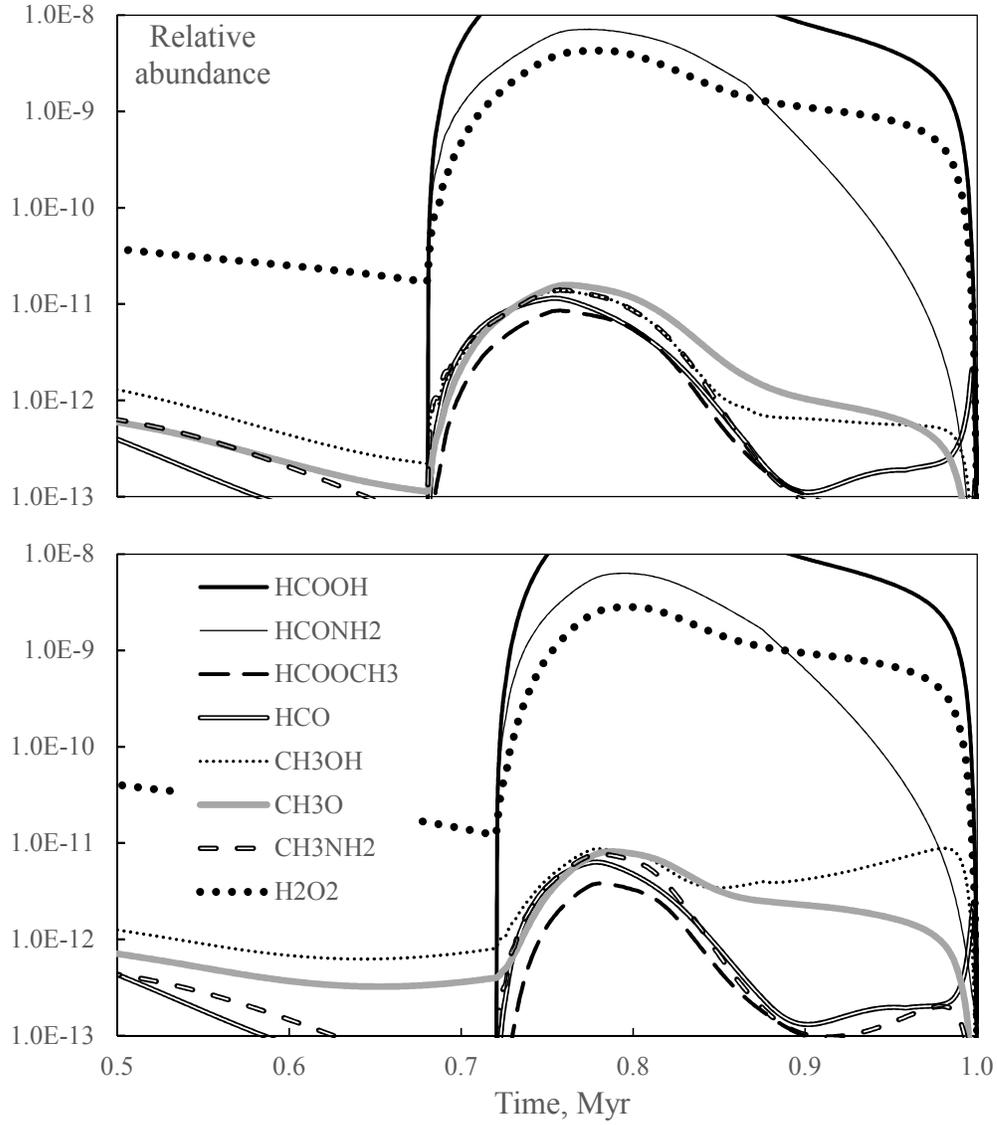}
 \vspace{-6cm}
 \caption{Gas-phase abundances, relative to hydrogen, of selected mantle-produced species for variants of Model K with $\alpha=0.01$ (top panel) and $\alpha=0.06$ (bottom panel).}
 \label{att-q-gascom}
\end{figure}
%
%
% Table 9
\begin{table*}
\begin{center}
\footnotesize
\caption{Comparison of observed and calculated (maximum) abundances of complex organic molecules in quiescent cloud cores.}
\label{tab-com}
\tabcolsep=0.11cm
    \begin{tabular}{lrlrlll}
\tableline
\tableline
 & \multicolumn{4}{c}{Observations, cm$^{-3}$} & \multicolumn{2}{c}{Model L, cm$^{-3}$} \\
Species & B1-b & & L1689b & & Gas\tablenotemark{1} & Ice\tablenotemark{2} \\
\tableline
H$_2$CO & 4.0E-10 & M\tablenotemark{3} & 1.3E-09 & B\tablenotemark{4} & 3.4E-10 & 1.2E-07 \\
HCO & 1.8E-11 & C\tablenotemark{5} & ... &  & 2.4E-10 & 2.2E-09 \\
HCOOH & 1.0E-11 & C & ... &  & 4.2E-08 & 1.4E-06 \\
HCOOCH$_3$ & 2.0E-11 & C & 7.4E-10 & B & 9.3E-13 & 7.7E-11 \\
CH$_3$OH & 3.1E-09 & O\tablenotemark{6} & ... &  & 1.4E-10 & 3.9E-09 \\
CH$_3$O & 4.7E-12 & C & ... &  & 2.4E-11 & 1.6E-12 \\
H$_2$CCO & 1.3E-11 & C & 2.0E-10 & B & 7.8E-12 & 9.2E-13 \\
CH$_3$CHO & 1.0E-11 & C & 1.7E-10 & B & 2.4E-13 & 1.3E-10 \\
HCCCHO & 3.6E-12 & C & ... &  & 8.2E-12 & 2.5E-12 \\
\tableline
\end{tabular}
\tablenotetext{1}{Maximum abundance}
\tablenotetext{2}{At $t=1$Myr}
\tablenotetext{3}{\citet{Marcelino05}}
\tablenotetext{4}{\citet{Bacmann12}}
\tablenotetext{5}{\citet{Cernicharo12}}
\tablenotetext{6}{\citet{Oberg10}}
\end{center}
\end{table*}
As stated in section~\ref{r-stand}, the model produces interesting results regarding complex organic molecules. COMs are efficiently produced in the sublayers by the photoprocess, and are released into the gas by photodesorption of species in shallow ice layers directly below the surface (section~\ref{gasgr}). Hydrogen-poor molecules, such as formic acid, formamide HCONH$_2$, formaldehyde CH$_2$O, glyoxal (CHO)$_2$, and other, heavier COMs may contain a significant part of oxygen and carbon reservoir. For the complete model, this produces a relatively high gas-phase abundances of many organic species between approximately 0.73 and 0.96Myr, or $A_V$ 2.4 to 10mag. The first number indicates the formation of first sublayers and the onset of subsurface PDI chemistry. The second number signals the end of the ISRF photons as a major source of desorption and dissociation, and the initiation of the rapid accretion phase, which prevents desorption of species whose formation time is longer that the time required to `bury' them in the mantle. Additionally, during the rapid freeze-out, thickness of the surface layer exceeds 2ML, at which point desorption from the sublayers is prohibited (section~\ref{gasgr}).
As stated in section~\ref{r-stand}, the model produces interesting results regarding complex organic molecules. COMs are efficiently produced in the sublayers by the photoprocess, and are released into the gas by photodesorption of species in shallow ice layers directly below the surface (section~\ref{gasgr}). Hydrogen-poor molecules, such as formic acid, formamide HCONH$_2$, formaldehyde CH$_2$O, glyoxal (CHO)$_2$, and other, heavier COMs may contain a significant part of oxygen and carbon reservoir. For the complete model, this produces a relatively high gas-phase abundances of many organic species between approximately 0.73 and 0.96Myr, or $A_V$ 2.4 to 10mag. The first number indicates the formation of first sublayers and the onset of subsurface PDI chemistry. The second number signals the end of the ISRF photons as a major source of desorption and dissociation, and the initiation of the rapid accretion phase, which prevents desorption of species whose formation time is longer that the time required to `bury' them in the mantle. Additionally, during the rapid freeze-out, thickness of the surface layer exceeds 2ML, at which point desorption from the sublayers is prohibited (section~\ref{gasgr}).

The reaction network has already been described in detail by \citet{Garrod06,Garrod08} and \citet{Laas11}. These authors investigate the formation of COMs during the gas warm-up phase in the protostellar stage. \citet{Vasyunin13b} investigate the gas-grain chemistry of COMs in cold cores. They use reactive desorption as the main means for heavy molecule ejection into the gas phase. They consider a clump dense gas (10$^5$cm$^{-3}$) existing in steady physical conditions for several hundred thousand years. The best-fit results obtained by \citep{Vasyunin13b} are for model with a very efficient reactive desorption process -- 10\% of reaction products go to the gas phase, and all ice species are affected by surface reactions (no bulk ice). Because of this, reactive desorption is attributed to all ice species and is summarily more effective roughly by a factor of $10^3$ than in Models J to M. \citet{Vasyunin13b} conclude that the assistance of other processes is required to reproduce gaseous COM abundances in protostellar cores.

The research presented here has shown that reactive desorption is very ineffective in the case of COMs. Meanwhile, surface synthesis of COMs is made more productive, because the large molecules remain on the grains. Moreover, it has been shown that photoprocessing of thin ices at low $A_V$, followed by photodesorption from shallow subsurface layers, can be a plausible source for COMs in the gas phase. This process is efficient at much lower densities and shorter evolutionary time-scales than those considered by \citet{Vasyunin13b}.

Figure~\ref{att-r-com} shows the calculated Model L abundances, relative to H, for a selection species that have been observed in dark clouds. These species include methanol CH$_3$OH, formic acid HCOOH, propynal HC$_2$CHO, acetaldehyde CH$_3$CHO, methyl formate HCOOCH$_3$, ketene CH$_2$CO, dimethyl ether CH$_3$OCH$_3$, methoxy (CH$_3$O), and formyl (HCO) radicals (see Table~\ref{tab-com}). All the observed abundances fall in the vicinity of $10^{-10} - 10^{-11}$, and the Standard model produced abundances in or near this range. The changes in desorption efficiencies in the complete model result in a poorer agreement with observations for several of the species.

In the following paragraphs a short analysis on the Model L chemistry for significant organic compounds, observed in quiescent cores, is presented. Table~\ref{tab-com} shows a comparison of observational and calculated abundances of organic species in dark clouds. The conditions and history of the regions sampled by observations are unknown. An agreement with observations can be claimed in cases, where the calculated maximum abundance equals or exceeds the observed value.

There is a dip for the ice abundance curves of radical species HCO and CH$_3$O, associated with the weakening of the ISRF. A similar behavior can be observed for most hydrogenated radical species. All radicals then experience a steep upward trend in the final stages, associated with the overall freeze-out of molecules.

\emph{Formaldehyde.} H$_2$CO is synthesized in ice in significant amounts (0.1\% of H$_2$O at 1Myr) via barrierless double-hydrogenation of CO by atomic H, although other pathways exist. It can be photodissociated back to CO or HCO, or converted to HCO by its reaction with the abundant OH radical. Other surface binary reactions of H$_2$CO have relatively high energy barriers and are ineffective.

\emph{Formic acid.} A barrierless reaction HCO+OH is the main pathway for the formation of HCOOH. Thus, formic acid is the end product of a barrierless reaction chain that involves CO and the abundant radicals H and OH. As such, HCOOH is produced in vast amounts in the sublayers, where OH is readily available. In Model L, the abundance of formic acid in ice reaches 1\% relative to that of H$_2$O. This is a significantly better result than the unrealistic 15\% for the Standard model. HCOOH is overproduced in the Standard model because of the prolonged PDI period, when CO is converted into HCOOH and other compounds. In turn, this period is so long because of the inefficient desorption in that model.

\emph{Other species formed in the mantle.} Methyl formate HCOOCH$_3$ has a low ice abundance of $6\times10^{-7}$ relative to H$_2$O at 1Myr. Its production requires the protonation of formaldehyde -- a reaction with a barrier. The same holds true for dimethyl ether CH$_3$OCH$_3$, whose synthesis does not involve any major radical species and thus is even lower than that of other COMs. Acetaldehyde CH$_3$CHO is able to form via the CH$_3$+HCO reaction. The synthesis of these two radicals involves no barriers (similarly to methanol), and CH$_3$CHO reaches somewhat higher abundance than other complex species.

\emph{Surface species.} Several of the species in Table~\ref{tab-com} can be associated exclusively with surface or gas-surface processes. The formation of CH$_3$OH occurs via an interaction of gas and surface chemistry. The hydrogenation of formaldehyde is hampered by activation barriers, and most of CH$_3$OH forms via the CH$_3$+OH reaction. This is a barrierless pathway that requires the successive hydrogenation of C with three H atoms. The radicals C, CH, CH$_2$, and CH$_3$ can also be available from the accretion of gas-phase species. The resulting production rate in ice for CH$_3$OH is low. It can be significantly higher in protostellar envelopes, where reactions with barriers become important \citep{Garrod06}. The metoxy radical CH$_3$O almost exclusively originates from the photodissociation of methanol.

Ketene H$_2$CCO and propynal HCCCHO are also associated with surface chemistry. Their synthesis starts with radicals that accrete directly from the gas, e.g., H, C, O, and C$_2$. Although barrierless, the synthesis involves a number of steps, and has a limited efficiency. The abovementioned radicals are not abundant in the sublayers. Because of this, H$_2$CCO and HCCCHO do not reach particularly high abundances in ice. However, the gas-phase abundance of these compounds is relatively high because of their concentration on the surface. This is especially obvious for the early core evolution stages (figure~\ref{att-r-com}).

Formic acid and formamide are produced in PDIs in vast amounts. These species form from radicals that are readily available in subsurface ice -- OH, HCO, and NH$_2$. The overproduction largely arises from the fact that most current reaction networks, including the one employed here, are designed for surface chemistry. Reactions for organic species with radicals that are almost non-existent on surface but abundant in subsurface layers -- NH, NH$_2$, OH, HCO, and others -- have largely been omitted in the network. These radicals arise from the direct photodissociation of the most common ice constituents, and should have a tremendous importance in the synthesis of complex molecules. A similar conclusion on subsurface ice chemistry was reached in \citet{Kalvans10}. The importance of such radicals in mantle reactions has been recognized by \citet{Garrod13a}, who included hydrogen abstraction reactions by OH in his three-phase model.

\citet{Garrod13a} successfully modeled the abundances of organic molecules during the warm-up and eventual evaporation of the (circumstellar) ices in the envelope of a protostar. In such conditions, cosmic-ray-induced photons are found to be important for radical generation in ices, instead of ISRF photons.

The main result from the Standard model remains valid also in the complete model -- organic molecules can be produced in PDIs in significant amounts, and their non-thermal release into the gas can be the responsible for the gaseous presence of at least several organic compounds in dark clouds. The two other complete models -- J and K -- produce gas-phase COM abundances within the same order of magnitude.

Because of its peculiar selectivity, direct reactive desorption may alter COM abundances in gas and ice. This mechanism has a dual effect -- more efficient desorption even more delays the formation of subsurface ice and the synthesis of COMs in the sublayers. For example, if the value of $\alpha$ is taken 0.1 instead of 0.03 for Model~L, the first sublayer forms only at 0.855Myr instead of 0.730Myr. However, small molecules are much more effectively desorbed, and the overall proportion of COMs in ice is higher, once the first sublayer has been initiated.

Variations of Models J, K, and L with $\alpha$ values in the range of 0.01-0.06 result in abundance differences up to several tens of per cent for the organic species. Figure~\ref{att-q-gascom} shows an example of gas-phase abundances with two different $\alpha$ values for selected species that are produced mostly in subsurface ice. The behavior of individual species can be highly specific. A general trend is a slight increase for gas-phase abundances of the heavier COMs at lower $\alpha$ values. Reactive desorption is highly dependent on the number of reactions involved in the formation of each species, and on the value of the parameter $f_{\rm rd}$ for each particular reaction. This is unlike the derivations (Models B-D) of the Standard model, where a higher $\alpha$ meant higher abundances of COMs. For the complete model, other desorption mechanisms hamper the formation of the sublayers, and any increase in the efficiency of reactive desorption delay the formation of subsurface ice even more, thus delaying the photo-synthesis of COMs. The overall result is that, counter-intuitively, more effective reactive desorption produces lower gas-phase abundances of COMs and other subsurface species. This also holds true for the solid phase.

Because complex species have similar assumed $Y_{\rm ph}$, photodesorption has no significant effect on COM abundances. Derivations of the Standard model have negligible differences on the order of a few per cent at most (section~\ref{r-pdes}). The effect of photodesorption in the complete models is even lower because sublayer ices are exposed to the ISRF for a short time, only.

\subsubsection{Other ice components}
\label{r-com-other}
%
% Figure 10
\begin{figure}
 \vspace{-10cm}
  \hspace{-2cm}
  \includegraphics{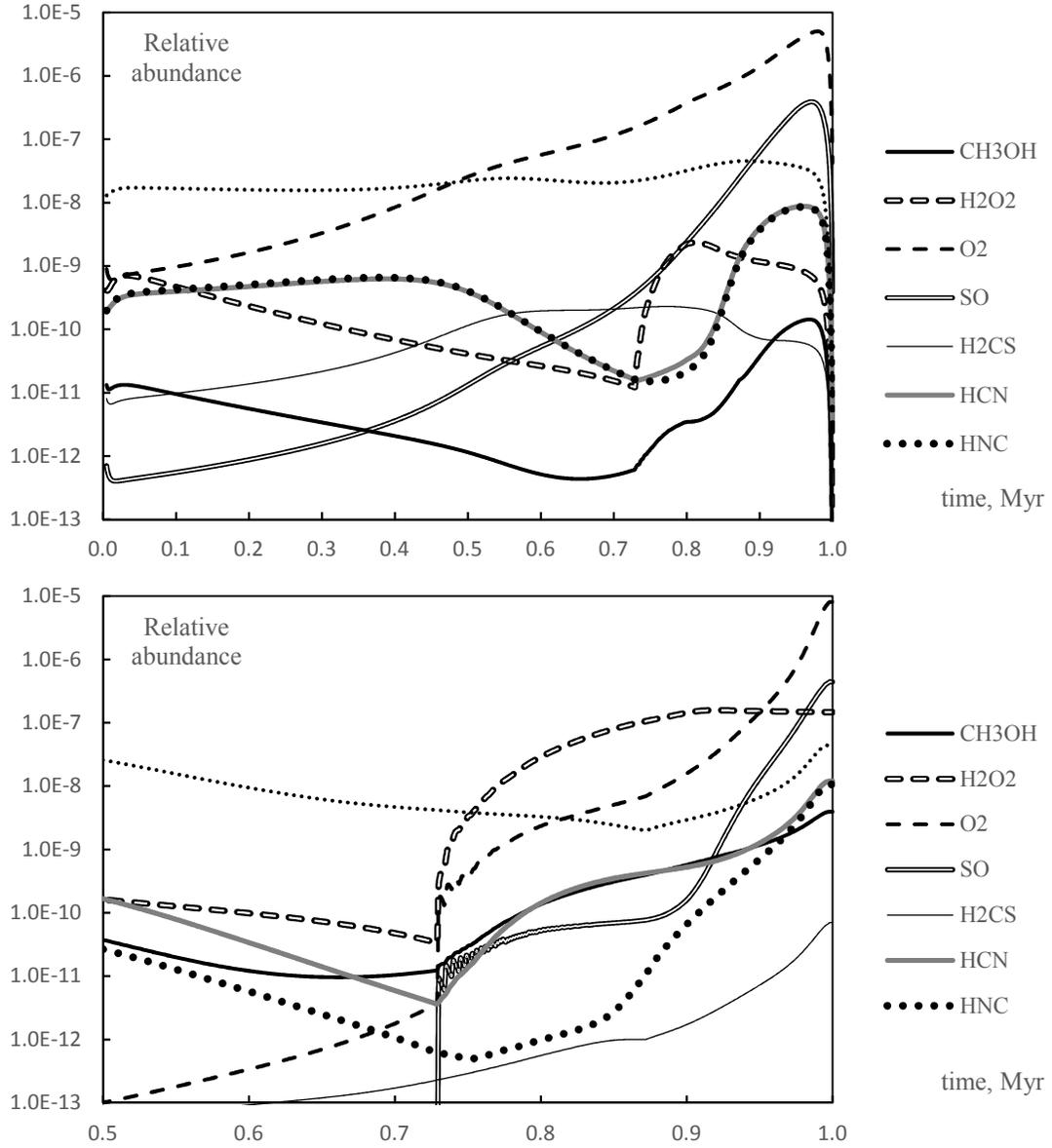}
 \vspace{-2cm}
 \caption{Gas-phase (top panel) and ice (bottom panel) abundances, relative to hydrogen, of the complete Model L for selected species.}
 \label{att-r-other}
\end{figure}
Figure~\ref{att-r-other} shows the abundance of selected species in the ice and the gas phase. Ice species can be broadly divided into two categories -- those produced on the surface (H$_2$O, CH$_3$OH, CO$_2$, H$_2$CO, H$_2$S), and those produced in the ice mantle (O$_2$, HCOOH, SO, SO$_2$, OCS, H$_2$O$_2$). During the late stages, a third category appears -- the molecules accreted directly from the gas phase (CO, N$_2$, HNC). Naturally, intermediate sub-categories exist with significant contributions from two or three categories (HCN, O$_2$).

Sulfur chemistry is dominated by surface production of H$_2$S. SO is the second most important sulfur molecule in ice with a final abundance of 42\% relative to that of H$_2$S. Oxides and OCS appear during late stages in the sublayers and reach abundances of 0.2 and 0.02\% relative to H$_2$S, respectively. Because OCS \citep{Palumbo97} and SO$_2$ \citep{Boogert97,Zasowski09} are the only solid sulfur compounds observed towards young stellar objects, a possible implication is that subsurface chemistry plays a significant role in grain surface processes \citep[see also][]{Kalvans10}.

The calculation results show an overabundance of solid ammonia, which is typically not observed in dark clouds \citep{Oberg11a}. The H$_2$O:NH$_3$ ice abundance ratio is 17\% for Model L and 20\% for the Standard model. Relevant models by other authors yield similar results \citep{Garrod11,Vasyunin13a}. The formation of NH$_3$ occurs simultaneously with H$_2$O, and via similar reactions. Because these two species have very similar $E_D$, they cannot be separated by selective desorption. A probable cause for such results is an underproduction of gas-phase N$_2$. This problem is only partially solved by the inclusion of N$_2$ shielding. A dedicated investigation of nitrogen gas-grain chemistry may be required to find a solution for this problem.  Ammonium ion NH$_4^+$ can be a major reservoir for nitrogen in ice, as suggested in previous studies \citep{Schutte03}.

Hydrogen peroxide has been observed in the interstellar medium with a high abundance of $10^{-10}$ relative to hydrogen \citep{Bergman11}. This has been attributed to surface chemistry by \citet{Du12}, with reactive desorption as the main mechanism. These authors also investigate in detail the formation mechanism of H$_2$O$_2$. This model was then verified by the observation of the O$_2$H radical with a similar abundance \citep{Parise12}.

In the complete model presented here, gaseous H$_2$O$_2$ is produced with relative abundances up to $10^{-9}$, in agreement with observations. Figure~\ref{r-com-other} shows a double production peak. H$_2$O$_2$ is produced by surface reactions, while subsequent hydrogenation gradually lowers its abundance. A second peak is observed with the onset of active mantle photochemistry. The model fails to produce gas-phase O$_2$H at relative abundances higher than $10^{-12}$. Differences between this model an the one employed by \citet{Du12} include the activation energy for the reaction H + O$_2$ (1200 and 600K, respectively) and different physical conditions. These likely do not produce significant changes for O$_2$H gas-phase abundance.

Most importantly, reactive desorption is more effective by a factor of $10^3$ in the models by \citet{Du12,Vasyunin13b}. This is because of their high $\alpha=0.1$ value, and because the hundred-monolayer-thick bulk mantle was not considered by these authors. The effectiveness of reactive desorption has to be carefully evaluated before application in two-phase models. It is possible that, if the model by \citet{Du12} considered subsurface mantle, then it could be difficult for this model to attain high gas abundances for species produced on surface.

The current paper certainly cannot offer a plausible explanation for the high O$_2$H abundance observed by \citet{Parise12}. The molecular cloud core in consideration, $\rho$ $Oph$ A is notable with observations of molecular oxygen and probably is in a transient evolutionary stage \citep{Liseau12}. A perturbed state is probably supported by the relatively high gas temperature of 20-30K, uncharacteristic for dark cores, and the existence of several protostars in the direct vicinity of the core \citep{Du12,Liseau12}.

\subsubsection{The composition of the sublayers}
\label{r-sub}
%
% Figure 11
\begin{figure}
 \vspace{-9cm}
  \hspace{1cm}
  \includegraphics[width=15.0cm]{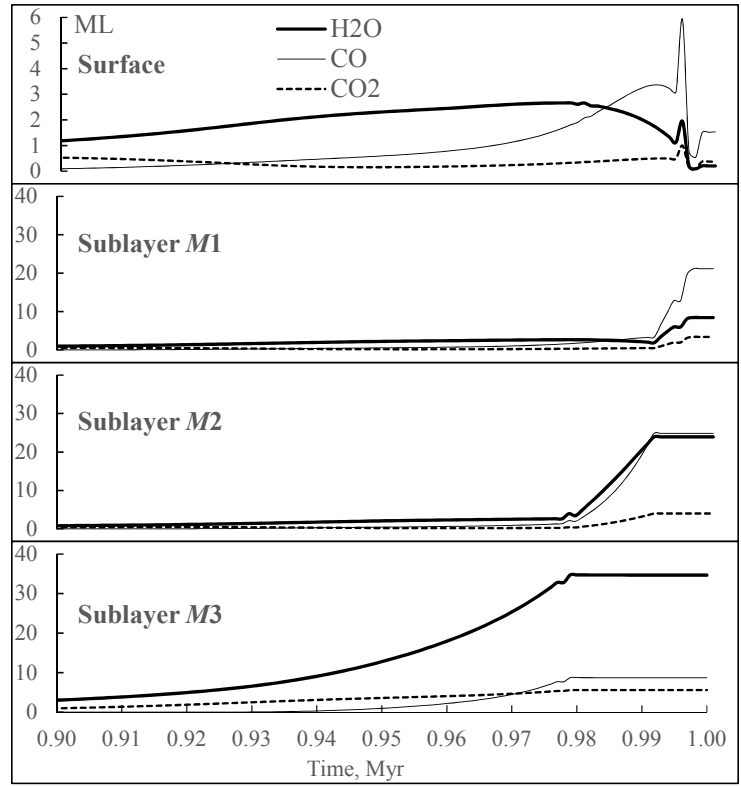}
%  \includegraphics{att-r-subab.eps}
% \vspace{-5cm}
 \caption{Model L results: abundances of water and carbon oxides in monolayers per sublayer, relative to total H, during the last 100kyr of cloud contraction.  The curved pattern for surface (upper panel) abundances is an artifact because of the surface-sublayer transition as described in section~\ref{comp}. Note the different abscissa scale for the surface layer. The composition of sublayers also reflects the evolution of surface composition over time.}
 \label{att-r-subab}
\end{figure}
In a fully formed ice at 1Myr, sublayer 3 has a H$_2$O:CO:CO$_2$:NH$_3$:N$_2$ ratio of 60:15:10:12:4, respectively. For sublayer 2 this ratio is 39:41:7:6:8, for sublayer 1 -- 22:54:7:2:13, and for the surface layer -- 7:50:12:1:21. The outer sublayer 1 is thinner -- $\approx$40ML instead of 60 for a `full' sublayer. Figure~\ref{att-r-subab} shows the evolution of water and carbon oxide abundances in the sublayers.  Sublayers 1 and 2 represent the two observed modes of solid CO -- in polar and in apolar ice matrix \citep{Sandford88,Tielens91}.
%This result removes the need to invoke rapid H$_2$O:CO ice segregation at $\approx$10K, as discussed by \citep{Garrod11}.

\citet{Bergin05} conclude that towards the field star Elias 16 ($A_V=24.1$) 15\% of CO$_2$ resides in apolar ice. For Model L at the corresponding $A_V$ the total ice thickness is 92.6ML, and only sublayer 3 is `full' with 60.6 MLs of ice in it. 74\% of CO$_2$ reside in sublayer 3, which is water-dominated (H$_2$O:CO 100:25), and the remainder in other layers, where H$_2$O:CO is roughly 1:1. These results probably are in an approximate agreement with observations. The composition of the surface layer indicates that the apolar CO most likely resides in the topmost layers if interstellar ices.

\section{Conclusions}
\label{concl}
% Table 10
\begin{table}
\begin{center}
\caption{List of desorption parameters employed in models.}
\label{tab-mod}
\begin{tabular}{llllc}
\tableline
\tableline
 & Photodes. & $\alpha$, reactive & H$_2$FD & H$_2$FD $E_D$  \\
Model & yield & desorption & efficiency & threshold (K) \\
\tableline
\small{Standard} & 0.001 & N/A\tablenotemark{1} & 0 & N/A \\
\tableline
A & uniform\tablenotemark{2} & St.\tablenotemark{3} & St. & N/A \\
\tableline
B & St. & 0.01 & St. & N/A \\
C & St. & 0.03 & St. & N/A \\
D & St. & 0.06 & St. & N/A \\
\tableline
E & St. & St. & $4\times10^{-6}$ & 1210 \\
F & St. & St. & $2\times10^{-6}$ & 2600 \\
G & St. & St. & $2\times10^{-6}$ & 6000 \\
H & St. & St. & RRK & 6000 \\
I & St. & St. & emp.\tablenotemark{4} & none \\
\tableline
J & uniform & 0.03 & $4\times10^{-6}$ & 2600 \\
K & uniform & 0.03 & RRK & 6000 \\
L & uniform & 0.03 & emp. & none \\
\tableline
M\tablenotemark{5} & uniform & 0.03 & RRK & 6000 \\
\tableline
\end{tabular}
\tablenotetext{1}{Reactive desorption efficiency assumed 1\%}
\tablenotetext{2}{0.002 for secondary and 0.003 for ISRF photons}
\tablenotetext{3}{As in Standard model}
\tablenotetext{4}{empiric $E_D$-dependent relation}
\tablenotetext{5}{Short (0.2Myr) core contraction time}
\end{center}
\end{table}
A three-phase model has been presented with several features that may aid in a better understanding of interstellar ice-related chemistry and physics. The concept of sublayers facilitates the modeling of chemical reactions in ice. This approach also has the potential to simulate protostellar ice evaporation in a complex way, as the sublayers are gradually exposed to the surface \citep[see][]{Fayolle11a}. This will be used in subsequent papers. The transition of molecules from the surface to the mantle with a finite rate allows the possibility to simulate the compaction of an initially porous ice (although this has little effect on ice composition). To a limited extent, desorption from the subsurface layers is permitted, which may help to explain the observed abundances of at least some organic molecules in dark clouds. Finally, the research suggests the existence of `photon-dominated ices' in weakly shielded regions of molecular clouds. Basically, shallow ice layers, isolated from the surface, may experience intensive processing by interstellar photons. This gives rise to a peculiar chemistry on grains during early stages of core contraction -- or in the exterior part of the cores.

The influence of three desorption mechanisms on ice composition (the H$_2$O:CO:CO$_2$ ratio) has been investigated in detail. Table~\ref{tab-mod} shows the summary of the specific models. K and L are the `final' models that yield the best agreement with observations for the $A_V$-dependent H$_2$O:CO:CO$_2$ ratio (section~\ref{r-compl}).

Photodesorption is fairly well constrained by laboratory results (Table~\ref{tab-pd}) and the adoption of $Y_{\rm ph}$ values for the ISRF photons that are in agreement with experiments is essential for a proper reproduction of ice accumulation in early core contraction stages. The values suggested here are $3\times10^{-3}$ for ISRF photons and $2\times10^{-3}$ for cosmic-ray induced photons, instead of a single value of $10^{-3}$ used in many current research papers \citep{Vasyunin13a,Vasyunin13b,Garrod13a,Gerner14,Chang14}.

For reactive desorption, it has sometimes been assumed that all binary reactions have similar ($\approx \alpha$) efficiencies \citep{Garrod07,Vasyunin13b}, while other authors use reaction- and molecule-specific desorption efficiencies \citep{Du12,Reboussin14}. The latter approach is the one used in the present study. No detailed analysis of selective reactive desorption could be found in the literature. The study of this mechanism in section~\ref{r-rdes} has yielded some unexpected results. Reactive desorption is inefficient for atom exchange reactions, characteristic for the formation of large molecules. These reactions typically do not produce much heat. Instead, the highly exothermic atom-addition reactions involved in the formation of simple species result in effective desorption for water, ammonia, and carbon dioxide.

Reactive desorption has a diverse effect on the abundances of COMs. Efficient desorption (higher $\alpha$) results in the accumulation of atoms in heavy molecules, because light species accumulate slower. This results in a thinner ice during early stages, and first sublayers appear later, which may significantly shorten the time available for organic synthesis in the sublayers. The latter effect is more important, because the effectiveness of PDI chemistry is strongly bound to ice thickness at low $A_V$ values. In any case, a reaction-specific approach on reactive desorption is essential for modeling of COM synthesis on interstellar dust grains.

The third mechanism in consideration was indirect reactive desorption, with desorption arising from the H+H reaction as its most important representation. It has an efficiency that is very poorly constrained by experiments or theory. It was found that this mechanism is probably responsible for the observed H$_2$O:CO:CO$_2$ ratio in interstellar ices. As a requirement for this, H$_2$FD has to be efficient for CO and CO$_2$, and negligible for water. Currently, as far as I know, there are no data that can either confirm or deny such a conclusion. The estimated efficiency parameter $f_{\rm H_2fd}$ likely has to be in the range $10^{-6}-10^{-4}$ desorbed molecules per accreted H atom for the CO molecule in order to produce a good agreement ($\pm$50\%) with observations. These $f_{\rm H_2fd}$ values do not counter the existing knowledge in the sense that they are two to four orders of magnitude lower that the efficiency of direct reactive desorption. Naturally, H$_2$FD is bound to the abundance of H atoms in the cloud, which might be a cause of uncertainty for $f_{\rm H_2fd}$. 

There are three general conclusions on ice accumulation. First, the onset of ice accumulation onto grains is mostly governed by desorption by interstellar photons, although direct and indirect reactive desorption also have an effect. Second, the observed $A_V$-dependent H$_2$O:CO:CO$_2$ ratio can be reproduced by a mechanism that is sufficiently effective and has a sharp difference in its yield for carbon oxides and water. Desorption by the H+H reaction on grains is a likely candidate. Third, efficient desorption during early stages of cloud contraction result in nominally slightly thicker ices in the dense cloud core. This is because the longer an ice layer is exposed to the ISRF, more multi-atom species are synthesized, and the total number of molecules is lower.

The three sublayers of subsurface ice represent three chemically distinct ice components at a stage when the freeze-out has ended. Starting from ice layer near the grain core itself, sublayer 3 consists of H$_2$O with an admixture of NH$_3$ and CO$_2$, sublayer 2 is a 1:1 mixture of H$_2$O and CO, and sublayer 1 is a 2:5 H$_2$O:CO mixture with an addition of N$_2$ and CO$_2$. The surface is largely covered with CO. These results are in compliance with observations of polar (water excess) and apolar CO and CO$_2$ ices (section~\ref{r-sub}).

As noted in section~\ref{r-2600}, all the models considered in this paper show an apparent depletion of CO ice at its respective $A_{\rm th}$ when compared to H$_2$O and CO$_2$. This result pertains also for the complete models (Table~\ref{tab-compl}). Because the $A_V$-dependent H$_2$O:CO:CO$_2$ ratio for Models K and L matches the observations, and because H$_2$O and CO$_2$ abundances at their respective $A_{\rm th}$ are quite similar, it can be possible to deduce a new detection threshold $A_V$ for solid CO. Based on these results, an estimate of $A_{\rm th}$ for CO is 10.5 from Model K and 8.1mag from Model L. Only the latter value lies within the error margin of the initially estimated $A_{\rm th}=6.8 \pm 1.6$ from observations by \citet{Bergin05}. Under the assumption that ice species have similar abundances at their respective $A_{\rm th}$, the model results suggest that the threshold $A_V$ for CO ice probably is in the range of 8--10.5mag, with the lower value being supported by currently published observations.

The higher spread for the calculated-to-observed CO ice abundance ratio (Table~\ref{tab-compl}) probably arises because, unlike water and CO$_2$, CO basically does not accumulate in ice until very late times. This means that a huge mass of CO accretes onto the grains on a very short time-scale, when the cloud has become sufficiently dark and dense. Consequently, relatively minor physical perturbations during late evolutionary stages of the core, caused by, e.g., nearby stars or outflows may speed-up or delay the CO, N$_2$, and O$_2$ accretion peak, producing the observed spread in abundances relative to H$_2$O, CO$_2$, and NH$_3$. The latter molecules can be expected to be less prone to such temporal perturbations because they begin to accumulate much earlier.

\acknowledgments
I acknowledge the support of Ventspils City Council. I would like to thank Dmitry Semenov for providing the code of `ALCHEMIC' astrochemical model \citep{Semenov10}, and the authors \citet*{Laas11} for providing their extensive reaction network. This research has made use of NASA's Astrophysics Data System.

\bibliography{mantle}

\begin{thebibliography}{}
\expandafter\ifx\csname natexlab\endcsname\relax\def\natexlab#1{#1}\fi

\bibitem[{{Accolla} {et~al.}(2011){Accolla}, {Congiu}, {Dulieu}, {Manic{\`o}},
  {Chaabouni}, {Matar}, {Mokrane}, {Lemaire}, \& {Pirronello}}]{Accolla11}
{Accolla}, M., {Congiu}, E., {Dulieu}, F., {et~al.} 2011, Physical Chemistry
  Chemical Physics (Incorporating Faraday Transactions), 13, 8037

\bibitem[{{Albertsson} {et~al.}(2013){Albertsson}, {Semenov}, {Vasyunin},
  {Henning}, \& {Herbst}}]{Albertsson13}
{Albertsson}, T., {Semenov}, D.~A., {Vasyunin}, A.~I., {Henning}, T., \&
  {Herbst}, E. 2013, \apjs, 207, 27

\bibitem[{{Andersson} {et~al.}(2006){Andersson}, {Al-Halabi}, {Kroes}, \& {van
  Dishoeck}}]{Andersson06}
{Andersson}, S., {Al-Halabi}, A., {Kroes}, G.-J., \& {van Dishoeck}, E.~F.
  2006, JChPh, 124, 064715

\bibitem[{{Andersson} \& {van Dishoeck}(2008)}]{Andersson08}
{Andersson}, S., \& {van Dishoeck}, E.~F. 2008, \aap, 491, 907

\bibitem[{{Awad} {et~al.}(2005){Awad}, {Chigai}, {Kimura}, {Shalabiea}, \&
  {Yamamoto}}]{Awad05}
{Awad}, Z., {Chigai}, T., {Kimura}, Y., {Shalabiea}, O.~M., \& {Yamamoto}, T.
  2005, \apj, 626, 262

\bibitem[{{Bacmann} {et~al.}(2012){Bacmann}, {Taquet}, {Faure}, {Kahane}, \&
  {Ceccarelli}}]{Bacmann12}
{Bacmann}, A., {Taquet}, V., {Faure}, A., {Kahane}, C., \& {Ceccarelli}, C.
  2012, \aap, 541, L12

\bibitem[{{Bahr} \& {Baragiola}(2012)}]{Bahr12}
{Bahr}, D.~A., \& {Baragiola}, R.~A. 2012, \apj, 761, 36

\bibitem[{{Belloche} {et~al.}(2009){Belloche}, {Garrod}, {M{\"u}ller},
  {Menten}, {Comito}, \& {Schilke}}]{Belloche09}
{Belloche}, A., {Garrod}, R.~T., {M{\"u}ller}, H.~S.~P., {et~al.} 2009, \aap,
  499, 215

\bibitem[{{Bergin} {et~al.}(2005){Bergin}, {Melnick}, {Gerakines}, {Neufeld},
  \& {Whittet}}]{Bergin05}
{Bergin}, E.~A., {Melnick}, G.~J., {Gerakines}, P.~A., {Neufeld}, D.~A., \&
  {Whittet}, D.~C.~B. 2005, \apjl, 627, L33

\bibitem[{{Bergman} {et~al.}(2011){Bergman}, {Parise}, {Liseau}, {Larsson},
  {Olofsson}, {Menten}, \& {G{\"u}sten}}]{Bergman11}
{Bergman}, P., {Parise}, B., {Liseau}, R., {et~al.} 2011, \aap, 531, L8

\bibitem[{{Bertin} {et~al.}(2013){Bertin}, {Fayolle}, {Romanzin}, {Poderoso},
  {Michaut}, {Philippe}, {Jeseck}, {{\"O}berg}, {Linnartz}, \&
  {Fillion}}]{Bertin13}
{Bertin}, M., {Fayolle}, E.~C., {Romanzin}, C., {et~al.} 2013, \apj, 779, 120

\bibitem[{{Boogert} \& {Ehrenfreund}(2004)}]{Boogert04}
{Boogert}, A.~C.~A., \& {Ehrenfreund}, P. 2004, in Astronomical Society of the
  Pacific Conference Series, Vol. 309, Astrophysics of Dust, ed. A.~N. {Witt},
  G.~C. {Clayton}, \& B.~T. {Draine}, 547

\bibitem[{{Boogert} {et~al.}(1997){Boogert}, {Schutte}, {Helmich}, {Tielens},
  \& {Wooden}}]{Boogert97}
{Boogert}, A.~C.~A., {Schutte}, W.~A., {Helmich}, F.~P., {Tielens},
  A.~G.~G.~M., \& {Wooden}, D.~H. 1997, \aap, 317, 929

\bibitem[{{Boogert} {et~al.}(2011){Boogert}, {Huard}, {Cook}, {Chiar}, {Knez},
  {Decin}, {Blake}, {Tielens}, \& {van Dishoeck}}]{Boogert11}
{Boogert}, A.~C.~A., {Huard}, T.~L., {Cook}, A.~M., {et~al.} 2011, \apj, 729,
  92

\bibitem[{{Brown} \& {Charnley}(1990)}]{Brown90}
{Brown}, P.~D., \& {Charnley}, S.~B. 1990, \mnras, 244, 432

\bibitem[{{Brown} {et~al.}(1988){Brown}, {Charnley}, \& {Millar}}]{Brown88}
{Brown}, P.~D., {Charnley}, S.~B., \& {Millar}, T.~J. 1988, \mnras, 231, 409

\bibitem[{{Caselli} {et~al.}(1998){Caselli}, {Hasegawa}, \&
  {Herbst}}]{Caselli98}
{Caselli}, P., {Hasegawa}, T.~I., \& {Herbst}, E. 1998, \apj, 495, 309

\bibitem[{{Cernicharo} {et~al.}(2012){Cernicharo}, {Marcelino}, {Roueff},
  {Gerin}, {Jim{\'e}nez-Escobar}, \& {Mu{\~n}oz Caro}}]{Cernicharo12}
{Cernicharo}, J., {Marcelino}, N., {Roueff}, E., {et~al.} 2012, \apjl, 759, L43

\bibitem[{{Chang} \& {Herbst}(2012)}]{Chang12}
{Chang}, Q., \& {Herbst}, E. 2012, \apj, 759, 147

\bibitem[{{Chang} \& {Herbst}(2014)}]{Chang14}
---. 2014, \apj, 787, 135

\bibitem[{{Cuppen} \& {Herbst}(2007)}]{Cuppen07}
{Cuppen}, H.~M., \& {Herbst}, E. 2007, \apj, 668, 294

\bibitem[{{Cuppen} {et~al.}(2009){Cuppen}, {van Dishoeck}, {Herbst}, \&
  {Tielens}}]{Cuppen09}
{Cuppen}, H.~M., {van Dishoeck}, E.~F., {Herbst}, E., \& {Tielens}, A.~G.~G.~M.
  2009, \aap, 508, 275

\bibitem[{{Du} {et~al.}(2012){Du}, {Parise}, \& {Bergman}}]{Du12}
{Du}, F., {Parise}, B., \& {Bergman}, P. 2012, \aap, 538, A91

\bibitem[{{Duley} {et~al.}(1989){Duley}, {Jones}, {Whittet}, \&
  {Williams}}]{Duley89}
{Duley}, W.~W., {Jones}, A.~P., {Whittet}, D.~C.~B., \& {Williams}, D.~A. 1989,
  \mnras, 241, 697

\bibitem[{{Duley} \& {Williams}(1988)}]{Duley88}
{Duley}, W.~W., \& {Williams}, D.~A. 1988, \mnras, 231, 969

\bibitem[{{Duley} \& {Williams}(1993)}]{Duley93}
---. 1993, \mnras, 260, 37

\bibitem[{{Fayolle} {et~al.}(2011b){Fayolle}, {Bertin}, {Romanzin}, {Michaut},
  {{\"O}berg}, {Linnartz}, \& {Fillion}}]{Fayolle11b}
{Fayolle}, E.~C., {Bertin}, M., {Romanzin}, C., {et~al.} 2011b, \apjl, 739, L36

\bibitem[{{Fayolle} {et~al.}(2011a){Fayolle}, {{\"O}berg}, {Cuppen}, {Visser},
  \& {Linnartz}}]{Fayolle11a}
{Fayolle}, E.~C., {{\"O}berg}, K.~I., {Cuppen}, H.~M., {Visser}, R., \&
  {Linnartz}, H. 2011a, \aap, 529, A74

\bibitem[{{Fayolle} {et~al.}(2013){Fayolle}, {Bertin}, {Romanzin}, {Poderoso},
  {Philippe}, {Michaut}, {Jeseck}, {Linnartz}, {{\"O}berg}, \&
  {Fillion}}]{Fayolle13}
{Fayolle}, E.~C., {Bertin}, M., {Romanzin}, C., {et~al.} 2013, \aap, 556, A122

\bibitem[{{Garrod} {et~al.}(2006){Garrod}, {Park}, {Caselli}, \&
  {Herbst}}]{Garrod06a}
{Garrod}, R., {Park}, I.~H., {Caselli}, P., \& {Herbst}, E. 2006, Faraday
  Discussions, 133, 51

\bibitem[{{Garrod}(2013{\natexlab{a}})}]{Garrod13a}
{Garrod}, R.~T. 2013{\natexlab{a}}, \apj, 765, 60

\bibitem[{{Garrod}(2013{\natexlab{b}})}]{Garrod13b}
---. 2013{\natexlab{b}}, \apj, 778, 158

\bibitem[{{Garrod} \& {Herbst}(2006)}]{Garrod06}
{Garrod}, R.~T., \& {Herbst}, E. 2006, \aap, 457, 927

\bibitem[{{Garrod} \& {Pauly}(2011)}]{Garrod11}
{Garrod}, R.~T., \& {Pauly}, T. 2011, \apj, 735, 15

\bibitem[{{Garrod} {et~al.}(2007){Garrod}, {Wakelam}, \& {Herbst}}]{Garrod07}
{Garrod}, R.~T., {Wakelam}, V., \& {Herbst}, E. 2007, \aap, 467, 1103

\bibitem[{{Garrod} {et~al.}(2008){Garrod}, {Weaver}, \& {Herbst}}]{Garrod08}
{Garrod}, R.~T., {Weaver}, S.~L.~W., \& {Herbst}, E. 2008, \apj, 682, 283

\bibitem[{{Gerakines} {et~al.}(1996){Gerakines}, {Schutte}, \&
  {Ehrenfreund}}]{Gerakines96}
{Gerakines}, P.~A., {Schutte}, W.~A., \& {Ehrenfreund}, P. 1996, \aap, 312, 289

\bibitem[{{Gerner} {et~al.}(2014){Gerner}, {Beuther}, {Semenov}, {Linz},
  {Vasyunina}, {Bihr}, {Shirley}, \& {Henning}}]{Gerner14}
{Gerner}, T., {Beuther}, H., {Semenov}, D., {et~al.} 2014, \aap, 563, A97

\bibitem[{{Gibb} {et~al.}(2004){Gibb}, {Whittet}, {Boogert}, \&
  {Tielens}}]{Gibb04}
{Gibb}, E.~L., {Whittet}, D.~C.~B., {Boogert}, A.~C.~A., \& {Tielens},
  A.~G.~G.~M. 2004, \apjs, 151, 35

\bibitem[{{Goldsmith} \& {Li}(2005)}]{Goldsmith05}
{Goldsmith}, P.~F., \& {Li}, D. 2005, \apj, 622, 938

\bibitem[{{Hartquist} \& {Williams}(1990)}]{Hartquist90}
{Hartquist}, T.~W., \& {Williams}, D.~A. 1990, \mnras, 247, 343

\bibitem[{{Hasegawa} \& {Herbst}(1993{\natexlab{a}})}]{Hasegawa93a}
{Hasegawa}, T.~I., \& {Herbst}, E. 1993{\natexlab{a}}, \mnras, 261, 83

\bibitem[{{Hasegawa} \& {Herbst}(1993{\natexlab{b}})}]{Hasegawa93b}
---. 1993{\natexlab{b}}, \mnras, 263, 589

\bibitem[{{Hasegawa} {et~al.}(1992){Hasegawa}, {Herbst}, \&
  {Leung}}]{Hasegawa92}
{Hasegawa}, T.~I., {Herbst}, E., \& {Leung}, C.~M. 1992, \apjs, 82, 167

\bibitem[{{Kalv{\= a}ns}(2013)}]{Kalvans13b}
{Kalv{\= a}ns}, J. 2013, Space Research Review, 2, 15

\bibitem[{{Kalv{\= a}ns} \& {Shmeld}(2010)}]{Kalvans10}
{Kalv{\= a}ns}, J., \& {Shmeld}, I. 2010, \aap, 521, A37

\bibitem[{{Kalv{\= a}ns} \& {Shmeld}(2013)}]{Kalvans13a}
---. 2013, \aap, 554, A111

\bibitem[{{Katz} {et~al.}(1999){Katz}, {Furman}, {Biham}, {Pirronello}, \&
  {Vidali}}]{Katz99}
{Katz}, N., {Furman}, I., {Biham}, O., {Pirronello}, V., \& {Vidali}, G. 1999,
  \apj, 522, 305

\bibitem[{{Laas} {et~al.}(2011){Laas}, {Garrod}, {Herbst}, \& {Widicus
  Weaver}}]{Laas11}
{Laas}, J.~C., {Garrod}, R.~T., {Herbst}, E., \& {Widicus Weaver}, S.~L. 2011,
  \apj, 728, 71

\bibitem[{{Lee} \& {Myers}(1999)}]{Lee99}
{Lee}, C.~W., \& {Myers}, P.~C. 1999, \apjs, 123, 233

\bibitem[{{Lee} {et~al.}(1996){Lee}, {Herbst}, {Pineau des Forets}, {Roueff},
  \& {Le Bourlot}}]{Lee96}
{Lee}, H.-H., {Herbst}, E., {Pineau des Forets}, G., {Roueff}, E., \& {Le
  Bourlot}, J. 1996, \aap, 311, 690

\bibitem[{{Leger}(1983)}]{Leger83}
{Leger}, A. 1983, \aap, 123, 271

\bibitem[{{Leger} {et~al.}(1985){Leger}, {Jura}, \& {Omont}}]{Leger85}
{Leger}, A., {Jura}, M., \& {Omont}, A. 1985, \aap, 144, 147

\bibitem[{{Li} {et~al.}(2013){Li}, {Heays}, {Visser}, {Ubachs}, {Lewis},
  {Gibson}, \& {van Dishoeck}}]{Li13}
{Li}, X., {Heays}, A.~N., {Visser}, R., {et~al.} 2013, \aap, 555, A14

\bibitem[{{Linnartz} {et~al.}(2011){Linnartz}, {Bossa}, {Bouwman}, {Cuppen},
  {Cuylle}, {van Dishoeck}, {Fayolle}, {Fedoseev}, {Fuchs}, {Ioppolo},
  {Isokoski}, {Lamberts}, {{\"O}berg}, {Romanzin}, {Tenenbaum}, \&
  {Zhen}}]{Linnartz11}
{Linnartz}, H., {Bossa}, J.-B., {Bouwman}, J., {et~al.} 2011, in IAU Symposium,
  Vol. 280, IAU Symposium, ed. J.~{Cernicharo} \& R.~{Bachiller}, 390--404

\bibitem[{{Liseau} {et~al.}(2012){Liseau}, {Goldsmith}, {Larsson}, {Pagani},
  {Bergman}, {Le Bourlot}, {Bell}, {Benz}, {Bergin}, {Bjerkeli}, {Black},
  {Bruderer}, {Caselli}, {Caux}, {Chen}, {de Luca}, {Encrenaz}, {Falgarone},
  {Gerin}, {Goicoechea}, {Hjalmarson}, {Hollenbach}, {Justtanont}, {Kaufman},
  {Le Petit}, {Li}, {Lis}, {Melnick}, {Nagy}, {Olofsson}, {Olofsson}, {Roueff},
  {Sandqvist}, {Snell}, {van der Tak}, {van Dishoeck}, {Vastel}, {Viti}, \&
  {Y{\i}ld{\i}z}}]{Liseau12}
{Liseau}, R., {Goldsmith}, P.~F., {Larsson}, B., {et~al.} 2012, \aap, 541, A73

\bibitem[{{Marcelino} {et~al.}(2005){Marcelino}, {Cernicharo}, {Roueff},
  {Gerin}, \& {Mauersberger}}]{Marcelino05}
{Marcelino}, N., {Cernicharo}, J., {Roueff}, E., {Gerin}, M., \&
  {Mauersberger}, R. 2005, \apj, 620, 308

\bibitem[{{Nejad} {et~al.}(1990){Nejad}, {Williams}, \& {Charnley}}]{Nejad90}
{Nejad}, L.~A.~M., {Williams}, D.~A., \& {Charnley}, S.~B. 1990, \mnras, 246,
  183

\bibitem[{{Noble} {et~al.}(2013){Noble}, {Theule}, {Borget}, {Danger},
  {Chomat}, {Duvernay}, {Mispelaer}, \& {Chiavassa}}]{Noble13}
{Noble}, J.~A., {Theule}, P., {Borget}, F., {et~al.} 2013, \mnras, 428, 3262

\bibitem[{{Oba} {et~al.}(2009){Oba}, {Miyauchi}, {Hidaka}, {Chigai},
  {Watanabe}, \& {Kouchi}}]{Oba09}
{Oba}, Y., {Miyauchi}, N., {Hidaka}, H., {et~al.} 2009, \apj, 701, 464

\bibitem[{{{\"O}berg} {et~al.}(2011{\natexlab{a}}){{\"O}berg}, {Boogert},
  {Pontoppidan}, {van den Broek}, {van Dishoeck}, {Bottinelli}, {Blake}, \&
  {Evans}}]{Oberg11b}
{{\"O}berg}, K.~I., {Boogert}, A.~C.~A., {Pontoppidan}, K.~M., {et~al.}
  2011{\natexlab{a}}, \apj, 740, 109

\bibitem[{{{\"O}berg} {et~al.}(2010){{\"O}berg}, {Bottinelli}, {J{\o}rgensen},
  \& {van Dishoeck}}]{Oberg10}
{{\"O}berg}, K.~I., {Bottinelli}, S., {J{\o}rgensen}, J.~K., \& {van Dishoeck},
  E.~F. 2010, \apj, 716, 825

\bibitem[{{{\"O}berg} {et~al.}({2009c}){{\"O}berg}, {Fayolle}, {Cuppen}, {van
  Dishoeck}, \& {Linnartz}}]{Oberg09c}
{{\"O}berg}, K.~I., {Fayolle}, E.~C., {Cuppen}, H.~M., {van Dishoeck}, E.~F.,
  \& {Linnartz}, H. {2009c}, \aap, 505, 183

\bibitem[{{{\"O}berg} {et~al.}(2011{\natexlab{b}}){{\"O}berg}, {Garrod}, {van
  Dishoeck}, \& {Linnartz}}]{Oberg11a}
{{\"O}berg}, K.~I., {Garrod}, R.~T., {van Dishoeck}, E.~F., \& {Linnartz}, H.
  2011{\natexlab{b}}, in 2010 NASA Laboratory Astrophysics Workshop, C44

\bibitem[{{{\"O}berg} {et~al.}({2009a}){{\"O}berg}, {Linnartz}, {Visser}, \&
  {van Dishoeck}}]{Oberg09a}
{{\"O}berg}, K.~I., {Linnartz}, H., {Visser}, R., \& {van Dishoeck}, E.~F.
  {2009a}, \apj, 693, 1209

\bibitem[{{{\"O}berg} {et~al.}({2009b}){{\"O}berg}, {van Dishoeck}, \&
  {Linnartz}}]{Oberg09b}
{{\"O}berg}, K.~I., {van Dishoeck}, E.~F., \& {Linnartz}, H. {2009b}, \aap,
  496, 281

\bibitem[{{Palumbo}(2006)}]{Palumbo06}
{Palumbo}, M.~E. 2006, \aap, 453, 903

\bibitem[{{Palumbo} {et~al.}(2010){Palumbo}, {Baratta}, {Leto}, \&
  {Strazzulla}}]{Palumbo10}
{Palumbo}, M.~E., {Baratta}, G.~A., {Leto}, G., \& {Strazzulla}, G. 2010,
  Journal of Molecular Structure, 972, 64

\bibitem[{{Palumbo} {et~al.}(1997){Palumbo}, {Geballe}, \&
  {Tielens}}]{Palumbo97}
{Palumbo}, M.~E., {Geballe}, T.~R., \& {Tielens}, A.~G.~G.~M. 1997, \apj, 479,
  839

\bibitem[{{Parise} {et~al.}(2012){Parise}, {Bergman}, \& {Du}}]{Parise12}
{Parise}, B., {Bergman}, P., \& {Du}, F. 2012, \aap, 541, L11

\bibitem[{{Pickles} \& {Williams}(1977)}]{Pickles77}
{Pickles}, J.~B., \& {Williams}, D.~A. 1977, ApSS, 52, 443

\bibitem[{{Prasad} \& {Tarafdar}(1983)}]{Prasad83}
{Prasad}, S.~S., \& {Tarafdar}, S.~P. 1983, \apj, 267, 603

\bibitem[{{Raut} {et~al.}(2007{\natexlab{a}}){Raut}, {Fam{\'a}}, {Teolis}, \&
  {Baragiola}}]{Raut07b}
{Raut}, U., {Fam{\'a}}, M., {Teolis}, B.~D., \& {Baragiola}, R.~A.
  2007{\natexlab{a}}, \jcp, 127, 204713

\bibitem[{{Raut} {et~al.}(2007{\natexlab{b}}){Raut}, {Teolis}, {Loeffler},
  {Vidal}, {Fam{\'a}}, \& {Baragiola}}]{Raut07a}
{Raut}, U., {Teolis}, B.~D., {Loeffler}, M.~J., {et~al.} 2007{\natexlab{b}},
  \jcp, 126, 244511

\bibitem[{{Reboussin} {et~al.}(2014){Reboussin}, {Wakelam}, {Guilloteau}, \&
  {Hersant}}]{Reboussin14}
{Reboussin}, L., {Wakelam}, V., {Guilloteau}, S., \& {Hersant}, F. 2014,
  \mnras, 440, 3557

\bibitem[{{Roberts} {et~al.}(2007){Roberts}, {Rawlings}, {Viti}, \&
  {Williams}}]{Roberts07}
{Roberts}, J.~F., {Rawlings}, J.~M.~C., {Viti}, S., \& {Williams}, D.~A. 2007,
  \mnras, 382, 733

\bibitem[{{Roser} {et~al.}(2001){Roser}, {Vidali}, {Manic{\`o}}, \&
  {Pirronello}}]{Roser01}
{Roser}, J.~E., {Vidali}, G., {Manic{\`o}}, G., \& {Pirronello}, V. 2001,
  \apjl, 555, L61

\bibitem[{{Ruffle} \& {Herbst}(2001b)}]{Ruffle01b}
{Ruffle}, D.~P., \& {Herbst}, E. 2001b, \mnras, 324, 1054

\bibitem[{{Sandford} {et~al.}(1988){Sandford}, {Allamandola}, {Tielens}, \&
  {Valero}}]{Sandford88}
{Sandford}, S.~A., {Allamandola}, L.~J., {Tielens}, A.~G.~G.~M., \& {Valero},
  G.~J. 1988, \apj, 329, 498

\bibitem[{{Schutte} \& {Khanna}(2003)}]{Schutte03}
{Schutte}, W.~A., \& {Khanna}, R.~K. 2003, \aap, 398, 1049

\bibitem[{{Semenov} {et~al.}(2010){Semenov}, {Hersant}, {Wakelam}, {Dutrey},
  {Chapillon}, {Guilloteau}, {Henning}, {Launhardt}, {Pi{\'e}tu}, \&
  {Schreyer}}]{Semenov10}
{Semenov}, D., {Hersant}, F., {Wakelam}, V., {et~al.} 2010, \aap, 522, A42

\bibitem[{{Strauss} {et~al.}(1994){Strauss}, {Chen}, \& {Loong}}]{Strauss94}
{Strauss}, H.~L., {Chen}, Z., \& {Loong}, C.-K. 1994, JCP, 101, 7177

\bibitem[{{Taquet} {et~al.}(2012){Taquet}, {Ceccarelli}, \&
  {Kahane}}]{Taquet12}
{Taquet}, V., {Ceccarelli}, C., \& {Kahane}, C. 2012, \aap, 538, A42

\bibitem[{{Tielens}(2005)}]{Tielens05}
{Tielens}, A.~G.~G.~M. 2005, {The Physics and Chemistry of the Interstellar
  Medium} (Cambridge University Press)

\bibitem[{{Tielens} {et~al.}(1991){Tielens}, {Tokunaga}, {Geballe}, \&
  {Baas}}]{Tielens91}
{Tielens}, A.~G.~G.~M., {Tokunaga}, A.~T., {Geballe}, T.~R., \& {Baas}, F.
  1991, \apj, 381, 181

\bibitem[{{Tomasko} \& {Spitzer}(1968)}]{Tomasko68}
{Tomasko}, M.~G., \& {Spitzer}, L. 1968, AJS, 73, 37

\bibitem[{{Turner}(1998)}]{Turner98}
{Turner}, B.~E. 1998, \apj, 501, 731

\bibitem[{{Vasyunin} \& {Herbst}(2013a)}]{Vasyunin13a}
{Vasyunin}, A.~I., \& {Herbst}, E. 2013a, \apj, 762, 86

\bibitem[{{Vasyunin} \& {Herbst}(2013b)}]{Vasyunin13b}
---. 2013b, \apj, 769, 34

\bibitem[{{Vasyunin} {et~al.}(2009){Vasyunin}, {Semenov}, {Wiebe}, \&
  {Henning}}]{Vasyunin09}
{Vasyunin}, A.~I., {Semenov}, D.~A., {Wiebe}, D.~S., \& {Henning}, T. 2009,
  \apj, 691, 1459

\bibitem[{{Watson} \& {Salpeter}(1972a)}]{Watson72a}
{Watson}, W.~D., \& {Salpeter}, E.~E. 1972a, \apj, 174, 321

\bibitem[{{Watson} \& {Salpeter}(1972b)}]{Watson72b}
---. 1972b, \apj, 175, 659

\bibitem[{{Whittet} {et~al.}(2001){Whittet}, {Gerakines}, {Hough}, \&
  {Shenoy}}]{Whittet01}
{Whittet}, D.~C.~B., {Gerakines}, P.~A., {Hough}, J.~H., \& {Shenoy}, S.~S.
  2001, \apj, 547, 872

\bibitem[{{Whittet} {et~al.}(2007){Whittet}, {Shenoy}, {Bergin}, {Chiar},
  {Gerakines}, {Gibb}, {Melnick}, \& {Neufeld}}]{Whittet07}
{Whittet}, D.~C.~B., {Shenoy}, S.~S., {Bergin}, E.~A., {et~al.} 2007, \apj,
  655, 332

\bibitem[{{Willacy} \& {Williams}(1993)}]{Willacy93}
{Willacy}, K., \& {Williams}, D.~A. 1993, \mnras, 260, 635

\bibitem[{{Willacy} {et~al.}(1994){Willacy}, {Williams}, \&
  {Duley}}]{Willacy94}
{Willacy}, K., {Williams}, D.~A., \& {Duley}, W.~W. 1994, \mnras, 267, 949

\bibitem[{{Williams}(1968)}]{Williams68}
{Williams}, D.~A. 1968, \apj, 151, 935

\bibitem[{{Zasowski} {et~al.}(2009){Zasowski}, {Kemper}, {Watson}, {Furlan},
  {Bohac}, {Hull}, \& {Green}}]{Zasowski09}
{Zasowski}, G., {Kemper}, F., {Watson}, D.~M., {et~al.} 2009, \apj, 694, 459

\end{thebibliography}
\bibliographystyle{apj}

\end{document}